\documentclass[aps,prd,showpacs,nofootinbib,11pt]{revtex4-1}
\usepackage[utf8]{inputenc}
\usepackage{amsmath,amsfonts,amssymb,graphicx,accents,mathrsfs,mathtools}
\usepackage{braket,textcomp}
\usepackage[percent]{overpic}
\usepackage{color}

\def\cosec{\text {cosec~}}
\def\sec{\text {sec}}
\begin{document}
\title{\boldmath An Inhomogeneous Jacobi equation for minimal surfaces and perturbative change in Holographic Entanglement Entropy}
\author{Avirup Ghosh}\email{avirup.ghosh@iitgn.ac.in}
\affiliation{Indian Institute of Technology, Gandhinagar, 382355, Gujarat , India.}
%and\\
%Theory Division, Saha Institute of Nuclear Physics, 1/AF Bidhan 
%Nagar, Kolkata 700064, India}
\author{Rohit Mishra,}\email{rohit.mishra@saha.ac.in}
\affiliation{Theory Division, Saha Institute of Nuclear Physics, HBNI, 1/AF Bidhan 
Nagar, Kolkata 700064, India}

\begin{abstract}
The change in Holographic entanglement entropy (HEE) for small fluctuations about pure anti-De Sitter ($AdS$) is  obtained by a perturbative expansion of the  area functional in terms of the change in the bulk metric and the embedded extremal surface. However it is known that change in the embedding appears  at second order or higher. It was shown that these changes in the embedding can be calculated in the 2+1 dimensional case by solving a ``generalized geodesic deviation equation''. We generalize this result to arbitrary dimensions by deriving an inhomogeneous form of the Jacobi  equation for minimal surfaces. The solutions of this equation map a minimal surface in a given space time to a minimal surface in a space time which is a perturbation over the initial space time. Using this we perturbatively calculate the changes in HEE upto second order for boosted black brane like perturbations over $AdS_4$.
\end{abstract}
\maketitle
\section{Introduction}
The Anti de Sitter $(AdS)$/Conformal Field Theory (CFT) correspondence \cite{Malda,Gubser,Witten} asserts that certain quantities like correlation functions of fields of a CFT living on the conformal boundary of $AdS$ can be obtained by calculating purely geometrical quantities in the higher dimensional bulk spacetime, which is a solution of a classical theory of gravity. One such quantity of interest is the entanglement entropy of a subregion $A$ in the boundary CFT. Following a proposal by Ryu and Takayanagi (RT)\cite{Ryu1,Ryu2} and later by Hubeny, Rangamani and Takayanagi (HRT)\cite{Hubeny}, this quantity can be holographically obtained by calculating the area of a spacelike co-dimension two `extremal surface' $(\gamma_A)$ in the bulk spacetime,
\begin{gather}\label{HEE}
 S_A ={Area(\gamma_A)\over 4 G_N},
\end{gather}
(where $G_N$ is the Newton's constant) and is dubbed as the Holographic Entanglement Entropy (HEE).  By extremal surface one refers to the following notion. For asymptotically $AdS$ spacetimes in $d+1$ dimensions the surface $\gamma_A$ is $(d-1)$ dimensional and is obtained by extremizing  the area functional,
\begin{gather}
 Area=\int d^{d-1}\sigma\sqrt{h},
\end{gather}
where $\sigma$'s are the intrinsic coordinates and $h_{ab}$ is the induced metric.

If this were also a minimum of the area functional, which is the case that arises in stationary and static geometries, then, according to the Holographic entanglement entropy  literature, it would be called a minimal surface. For static geometries the timelike Killing vector ($\partial_t$ say) is hypersurface orthogonal in the bulk geometry. It can then be shown that the extremal surface must lie on $t=constant$ slice and can be shown to be minimal. Hence the proposal reduces to finding a minimal surface on a constant time slice. The proposal, initially put forward by RT was precisely this. However for non static cases, where the timelike killing vector is not hypersurface orthogonal, or for dynamical geometries, where there is no time like Killing vector, $\gamma_A$ is no more minimal, and therefore RT proposal fails and one has to resort to the more general HRT proposal. {\it(In terms of nomenclature, in the mathematics literature, a minimal surface refers to just the critical point of the area functional and may not correspond to the minimum of the functional \cite{anciaux}. This is particularly the case in manifolds endowed with a Semi-Riemannian metric. We will stick to the latter nomenclature and use extremal and minimal interchangeably. Hence when we say minimal surfaces we actually mean extremal surfaces of HRT)} The equation obtained by extremizing the functional turns out to be  nothing but the condition that the trace of the extrinsic curvature of the surface vanishes. The condition however yields non linear equations of motion for the embedding functions. It therefore becomes difficult to solve these equations unless the back ground geometry is highly symmetric. Consequently, though these equations for the embedding function can be obtained exactly for $AdS$ it becomes difficult to solve them exactly even for backgrounds like the boosted black brane or the Kerr-AdS. One therefore considers doing a perturbation by treating these backgrounds as perturbations over $AdS$, near the asymptotic boundary. This imminently yields linear equations as the procedure involves a linearization of the minimal surface equation. 

The change in HEE between $AdS$ and excitations over it can then be calculated by considering variation of the area functional which incorporates the changes due to the change in the extremal surface $\gamma_A$ and the perturbation of the bulk metric.
At first order contributions only come from metric perturbations alone, while the change of the embedding of the extremal surface does not \cite{Lashkari,Jyotirmoy,Nozaki}. However at second order both first order change in the embeddings and second order metric perturbations contribute \cite{He, He:2013rsa,Guo:2013aca,Lashkari:2015hha, Kim:2015rvu}. In a previous work \cite{Ghosh:2016fop} the authors proposed a way to calculate the contributions to second order variations coming from the changes in the embedding, in $2+1$ dimensions. This was achieved by studying geodesic deviations between geodesics in rotating BTZ black hole (seen as perturbation over pure $AdS$) and pure $AdS_3$. These deviations were obtained as solutions of a ``generalized geodesic deviation equation''. In this paper we shall generalize this to arbitrary dimensions. In order to do so one has to reproduce the above notion, but now for minimal surfaces. Simplified cases for this deformation problem can be found in \cite{Fursaev:2010ix}.

Study of minimal surfaces in Riemannian geometries has been extensively carried out in the mathematics literature \cite{minimal, anciaux}. In the entanglement entropy literature the plateau problem for minimal surfaces has been studied in \cite{Fursaev:2007sg}. It is known that for surfaces embedded in a a given Riemannian space the area functional of the embedded surface is stationary, that is it's first variation vanishes, when the embedded surface is minimal. Likewise when the second variation is equated to zero it gives rise to the Jacobi equation for minimal surfaces \cite{simons}. The interpretation of the solutions of the Jacobi equation is the following. The solutions of this equation gives the deviation between a minimal surface and a neighboring minimal surface. In the physics literature the Jacobi equation has been studied in the context of relativistic membranes \cite{Capovilla:1994bs} and spiky strings on a flat background \cite{Bhattacharya:2016ixc}. However this equation is relevant only when the metric of the ambient space is fixed. 

In the context of the present work one needs to modify this notion. Note that in our case one needs to study deviations between two surfaces which are minimal in two different spacetimes. The spacetimes are however related by a perturbation and not completely arbitrary. To begin with one has to ensure that all of the results obtained are manifestly gauge invariant and therefore has to be careful and precise in defining perturbations in the spirit of a covariant perturbation theory. We therefore adopt the notion introduced in \cite{Stewart:1974uz} in the context of gravity. A priori, taking cue from the results obtained for geodesics one then expects the Jacobi equation to be modified by appearance of an inhomogeneous term. This indeed turns out to be case, as will be shown later. We also obtain an expression for the change in the area functional, in arbitrary dimensions, upto second order.

Having obtained an equation that properly mimics the situtation at hand, one needs to demonstrate that the equations can indeed be solved, for the prescription to be of any relevance. We therefore solve this equation in the $3+1$ dimensional case for two choices of the boundary subsystem 1) Spherical subsystem and 2) Thin strip subsystem. We do this for  Boosted black brane like  perturbations over $AdS_4$. Using the solutions of the inhomogeneous Jacobi equation we obtain the change in HEE between $AdS_4$ and boosted black brane like perturbations over it.

\section{Notations and conventions}\label{N&C}
Consider a $d+1$ dimensional space time $(\mathcal M,{g})$ and another $d+1$ dimensional space time $(\mathcal M',g')$ which is diffeomorphic to $\mathcal M$. That is there is a differentiable map $\Phi:\mathcal M\rightarrow \mathcal M'$ which is however not isometric. We will call $(\mathcal M',g')$ to be a perturbation over $(\mathcal M,g)$ if $\accentset{(1)}{P}=\Phi_*g'-g$ is a small perturbation over $g$. Consider a surface $\mathcal S$ isometrically embedded in $\mathcal M$ and given by the function $f:S\rightarrow \mathcal M$. It is implied that the restriction of $f$ to the image of $\mathcal S$ is continuous and differentiable. In a local coordinate chart $x^\mu$ on $\mathcal M$ and $\tau^a$ on $\mathcal S$ the embedding can be represented by the embedding functions $x^\mu\circ f\circ (\tau^a)^{-1}$. This can be simply written as $x^{\mu}(\tau^a)$. The induced metric on $\mathcal S$ is the pull back of the metric $g$ under the map $f$, given by $h=f_*~g$. Again, in the local coordinates this can be written as $h_{ab}=g(\partial_a,\partial_b)=\frac{\partial x^\mu}{\partial\tau^a}\frac{\partial x^\nu}{\partial\tau^b}g(\partial_\mu,\partial_\nu)$. The quantity $\frac{\partial x^\mu}{\partial\tau^a}\partial_\mu$ is the push forward of the purely tangential vector field $\partial_a$ to $\mathcal M$. `$h_{ab}$' is the first fundamental form on $\mathcal S$. To define the second fundamental form one needs a connection or the covariant derivative on $\mathcal M$. The covariant derivative is a map $\nabla: T\mathcal M \otimes T\mathcal M\rightarrow T\mathcal M$. For two vector fields $W,Z~\in ~T\mathcal M$ it is denoted as $\nabla_WZ$ and is an element of $T\mathcal M$. Now suppose $x\in\mathcal S$. One can decompose the tangent space at the point $x$ into the tangent space of $\mathcal S$ and the space of normal vectors as $T_x\mathcal M=T_x\mathcal S\oplus T_x^{\perp}\mathcal S$. Then one defines the tangent bundle and normal bundle on $\mathcal S$ as $\bigcup_xT_x\mathcal S$ and $\bigcup_xT_x^\perp\mathcal S$ respectively. One can similarly define a covariant derivative on $\mathcal S$. Let it be denoted by $D: T\mathcal S \otimes T\mathcal S\rightarrow T\mathcal S$. Let $X,Y~\in T\mathcal S$. Then the Gauss decomposition allows us to write,
\begin{gather}
\nabla_XY=D_XY+K(X,Y),
\end{gather}
where $D_XY$ is purely tangential and $K(X,Y)$ is a vector in the normal bundle and is the extrinsic curvature or the second fundamental form. The metric compatibility of $\nabla$ in this notation is written as $\nabla_Wg(V,U)=g(\nabla_WU,V)+g(U,\nabla_WV)$. The metric compatibility of $\nabla$ with $g$ will imply metric compatibility of $D$ with $h$, by virtue of the above equation. One defines a connection $\nabla^{\perp}_XN^\perp$ in the normal bundle as $\nabla^\perp:T\mathcal S\otimes T^\perp \mathcal S\rightarrow T^\perp \mathcal S$, where $X\in T\mathcal S$ and $N^\perp\in T^\perp\mathcal S $. Then the shape operator $W_{N^\perp}(X)$ is defined as,
\begin{gather}
\nabla_XN^\perp=\nabla_X^\perp N^\perp-W_{N^\perp}(X).
\end{gather} 
The shape operator and the extrinsic curvatures are related by the Weingarten equation,
\begin{gather}
g(W_{N^\perp}(X),Y)=g(N^\perp,K(X,Y)),
\end{gather}
where $X,Y\in T\mathcal S$ and $N^\perp\in T^\perp\mathcal S$. The Riemann tensor is a map $R:T\mathcal M \otimes T\mathcal M \otimes T\mathcal M\rightarrow T\mathcal M$ and is defined as,
\begin{gather}
R(W,U)V\equiv[\nabla_W,\nabla_U]V-\nabla_{[W,U]}V
\end{gather}
Similarly one can define an intrinsic Riemann tensor by,
\begin{gather}
\mathcal R(X,Y)Z\equiv[D_X,D_Y]Z-D_{[X,Y]}Z
\end{gather}
We write down the equations of Gauss and Codazzi, in this notation. Let $X,Y,Z,W\in T\mathcal S$ and $N^\perp\in T^\perp\mathcal S$. Then the Gauss equation is given as,
\begin{gather}
g(R(X,Y)Z,W)=g(\mathcal R(X,Y)Z,W)-g(K(X,Z),K(Y,W))+g(K(X,W),K(Y,Z)),
\end{gather}
and the Codazzi equation as,
\begin{gather}
g(R(X,Y)N^\perp,Z)=g((\nabla_YK)(X,Z),N^\perp)-g((\nabla_XK)(Y,Z),N^\perp)
\end{gather}
Now, we go over to notations involving perturbations. In the presence of perturbations a variation will be assumed to have have two contributions, one which is a flow along a vector $N\in T\mathcal M$, obtained by taking a covariant derivative $\nabla_N$ along $N$ and another variation $\delta_g$ which is purely due to metric perturbations. Since we will be doing all the calculations in a coordinate chart in the unperturbed space time, let try to define certain quantities on $\mathcal M$ arising due to the perturbations, i.e due to the difference in the two metrics $g$ and $\Phi_*~g'$. The metric perturbation will be given by,
\begin{gather}
(\delta_g~g)(\partial_\mu,\partial_\nu)\equiv\bigg[\Phi_*~g'-{g}\bigg](\partial_\mu,\partial_\nu)=\accentset{(1)}{P}(\partial_\mu,\partial_\nu),
\end{gather}
where $\accentset{(1)}{P}$ is a symmetric bilinear form on $\mathcal M$. Note that $\delta_g$ only acts on the metric and does not change the vector fields $\partial_\mu$. Now suppose there is a covariant derivative $\nabla'$ in $\mathcal M'$ compatible with $g'$, then for $X,Y\in T\mathcal M$,
\begin{gather}
C(X,Y)\equiv\delta_g\bigg(\nabla_XY\bigg)=\tilde\nabla_XY-\nabla_XY,
\end{gather}
where $\tilde\nabla=\phi^*\nabla'$ is the pullback connection on $M$. Note that $C(X,Y)$ is a vector field in $\mathcal M$. When written in coordinates it has exactly the same form as $\accentset{(1)}{C}^\mu_{\nu\rho}$ used in \cite{Ghosh:2016fop}. Since we will not be dealing with perturbations of further higher order, we have dropped the superscript $\accentset{(1)}{~}$. 

We are now in a position to derive the inhomogeneous Jacobi equation for minimal surfaces. For the display of some semblance with \cite{Ghosh:2016fop}, a rederivation of the inhomogeneous Jacobi equation for geodesics, in this notation, is given in Appendix \ref{geosection}. 
%%%%%%%%%%%%%%%%%%%%%%%%%%%%%%%%%%%%%%%%%%%%%%%%%%%%%%%%%%%%%%%%%%%%%%%%%%%%%%%%%%%%%%%%%%%%%%%%%%%%%%%%%%%%%%%%%%%%%%%%%%%%%%%%%%%%%%%%%%%%%%%%%%%%%%%%%%%%%%%%%%%%%%%%%%%%%%%%%%%%%%%%%%%%%%%%%%%%%%%%%%%%%%%%%%%%%%%%%%%%%%%%%%%%%%%%%%%%%%%%%%%%%%%%%%%%%%%%%%%%%%%%%%%%%%%%%%%%%%%%%%%%%%%%%%%%%%%%%%%%%%%%
\section{Derivation of the Inhomogeneous Jacobi equation for surfaces}\label{DOJE}
In the previous section we considered $(\mathcal M',g')$ to be a perturbation over $(\mathcal M, g)$. Let us consider a one parameter family of such perturbed spacetimes $(\mathcal M_\lambda, g_\lambda)$ and a one parameter family of diffeomorphism, which are not necessarily isometric, $\Phi_\lambda:\mathcal M\rightarrow \mathcal M_\lambda$ such that $\mathcal M_0$ corresponds to the unperturbed spacetime and $\Phi_0$ is the identity map. Let $\mathcal S_\lambda$ be a family of co-dimension two minimal surfaces in $(\mathcal M_\lambda,g_\lambda)$ i.e the trace of their extrinsic curvatures vanishes. The surfaces can be parametrized by the embedding functions $f^\mu_\lambda (\tau^a)$, which allows one to write the tracelessness condition as $h^{ab}_\lambda K_{(\lambda)}(\partial_a,\partial_b)=0$. Note that one would think that the coordinates $\tau^a$ may be different for different $\mathcal S_\lambda$. But one can always adjust the functions $f^\mu_\lambda$ such that the surfaces can be coordinatised by the same intrinsic coordinates. Let us construct a family of immersed submanifolds $\tilde {\mathcal S_\lambda}$ in $\mathcal M_0$, given by the embedding functions $F^\mu_\lambda$ such that $\Phi_\lambda\circ F^\mu_\lambda=f^\mu_\lambda$. Let's denote the deviation vector between $F_0^\mu$ and the neighboring surface be denoted by $N$. Note that $N$ can always be taken to be normal to $\tilde{\mathcal {S}}_0$, as any tangent deviation will only result in a reparametrization of the intrinsic coordinates $\tau ^a$ and won't change the area of the surface. This statement is however not obvious in our case where we have metric perturbations. In this regard we take cue from the calculation done in the case of geodesic \cite{Ghosh:2016fop}. Since we have already removed the freedom of intrinsic coordinate reparametrization, by adjusting the $f_\lambda$'s, it is quite legitimate to take normal variations only. Moreover since we will ultimately be interested in area change it is sufficient for us to take normal variations only. Further $N$ can always be chosen such that it commutes with the vectors $\partial_a$ tangent to the submanifold i.e $[N,\partial_a]=0~\forall~a$.

The condition that $\mathcal S_\lambda$'s are minimal in $(\mathcal M_\lambda,g_\lambda)$ then reduces to a condition on $N$ in $\mathcal M_0$. At each order of the variation, the conditions are essentially inhomogeneous linear differential equations that $N$ must satisfy. The equation that one obtains at linear order is the one we will be interested in, since the solutions of this will provide us with the linear deformation of the minimal surface that we are seeking. As is evident, the equation can be derived by equating the more general variation $\delta_N=\nabla_N+\delta_g$, discussed in section \ref{N&C}, of the trace of the extrinsic curvature to zero i.e
\begin{equation}\label{bs}
 \delta_N H_\lambda=h^{ab}_\lambda(\delta_N (\nabla_{(\lambda)\partial_a}\partial_b)^\perp)+(\delta_Nh^{ab}_{\lambda})K_{\lambda}(\partial_a,\partial_b)=0.
\end{equation}
We will drop the $\lambda$ subscript from here on, as the above variations will be calculated around the unperturbed surface i.e at $\lambda=0$. While dropping the $\lambda$'s surely will make the expressions look cleaner, one has to make sure that the minimal surface equation be used only after the derivatives have been computed. Let us first compute the first term of the above expression which involves the normal component of the covariant derivative. 
\begin{eqnarray}\label{losq}
 h^{ab}\delta_N (\nabla_{\partial_a}\partial_b)^\perp
 =h^{ab}\Biggl(\nabla_N(\nabla_{\partial_a}\partial_b)+\delta_g(\nabla_{\partial_a}\partial_b)-\nabla_N(\nabla_{\partial_a}\partial_b)^T-\delta_g (\nabla_{\partial_a}\partial_b)^T\Biggr)\nonumber\\
=h^{ab}\Biggl(\nabla_{\partial_a}\nabla_{\partial_b}N+R(N,\partial_a)\partial_b+C(\partial_a,\partial_b)-\nabla_N(\nabla_{\partial_a}\partial_b)^T-\delta_g (\nabla_{\partial_a}\partial_b)^T\Biggr)
\end{eqnarray}
 The action of the variation $\delta_N$ on any quantity $Q$ on $\mathcal M_0$ is taken to be of the form $\delta_N(Q)=\nabla_N(Q)+\delta_g(Q)$. This notation for variation has been adopted for convenience of calculation. That this reproduces the correct result, can be seen from the derivation of the inhomogeneous Jacobi equation, obtained by adopting this notation (appendix \ref{geo}). The action of $\delta_g$ is precisely on the space of sections on a tensor bundle in $\mathcal M_\lambda$. If we represent  a flow on $\mathcal M_0$ and $\delta_g$ by two parameters then a priori these two parameters are completely independent of each other, but for the perturbations to work one needs them to be equal. How the parameter of the flow $\nabla_N$ can be related to the parameter of the variation $\delta_g$ is a mathematical issue the resolution of which we will leave for some future work. Adopting the above, one obtains,
\begin{equation}\label{cs}
 (\delta_Nh^{ab})K(\partial_a,\partial_b)=2h^{ab}K(\partial_a,W_{N}(\partial_b))-h^{ac}h^{bd}K(\partial_a,\partial_b)\accentset{(1)}{P}(\partial_c,\partial_d)
\end{equation}

Substituting \eqref{losq},\eqref{cs} in \eqref{bs} we get 
\begin{gather}\label{nst}
 \delta_N H=h^{ab}\Biggl(\nabla_{\partial_a}\nabla_{\partial_b}N+R(N,\partial_a)\partial_b+C(\partial_a,\partial_b)-\nabla_N(\nabla_{\partial_a}\partial_b)^T-\delta_P (\nabla_{\partial_a}\partial_b)^T\Biggr)\\\notag+2h^{ab}K(\partial_a,W_{N}(\partial_b))-h^{ac}h^{bd}K(\partial_a,\partial_b)\accentset{(1)}{P}(\partial_c,\partial_d).
\end{gather}
A similar exercise with the term $ h^{ab}\delta_N (\nabla_{\partial_a}\partial_b)^T$ yield the following expression,
\begin{gather}\label{ct}
h^{ab}\Biggl[(\nabla_{(\nabla_{\partial_a}\partial_b)^T}N)^{\perp}+(\nabla_{\partial_a}\nabla_{\partial_b}N+R(N,\partial_a)\partial_b+C(\partial_a,\partial_b))^T+h^{cd}\accentset{(1)}{P}(K(\partial_a,\partial_b),\partial_c)\partial_d\Biggr].
\end{gather}
Substituting \eqref{ct} in \eqref{nst}, we get a complete expression for $\delta_NH$,
\begin{gather}\label{gost}
 \delta_NH=h^{ab}\Biggl((\nabla_{\partial_a}\nabla_{\partial_b}N+R(N,\partial_a)\partial_b+C(\partial_a,\partial_b))^{\perp}-(\nabla_{(\nabla_{\partial_a}\partial_b)^T}N)^{\perp}\Biggr)-h^{cd}\accentset{(1)}{P}(H,\partial_c)\partial_d\\\notag
 +2h^{ab}K(\partial_a,W_{N}(\partial_b))-h^{ac}h^{bd}K(\partial_a,\partial_b)\accentset{(1)}{P}(\partial_c,\partial_d)
\end{gather}
Noting that $(\nabla_{\partial_a}\nabla_{\partial_b}N)^\perp=-K(\partial_a,W_{N}(\partial_b))+\nabla^{\perp}_{\partial_a}\nabla^{\perp}_{\partial_b}N$, the above equation, along with the minimality condition $H=0$, can be recast in the following form, which is closer in form to the expressions known in the literature of minimal surfaces.
\begin{gather}\label{bang}
  \delta_NH%=h^{ab}\Biggl(\nabla^{\perp}_{\partial_a}\nabla^{\perp}_{\partial_b}N-%\nabla^{\perp}_{(\nabla_{\partial_a}\partial_b)^T}N+(R(N,\partial_a)\partial_b)^{\perp}%+K(\partial_a,W_{N}(\partial_b))+(C(\partial_a,\partial_b))^{\perp}\Biggr)\\\notag-%h^{ac}h^{bd}K(\partial_a,\partial_b)P(\partial_c,\partial_d)\notag\\
  =\Delta^{\perp}N+Ric(N)+A(N)+C^\perp-\tilde{H},
\end{gather}
where we have defined  $\Delta^{\perp}N$ to be the Laplacian on the normal bundle, given by $h^{ab}\Bigl(\nabla^{\perp}_{\partial_a}\nabla^{\perp}_{\partial_b}N-\nabla^{\perp}_{(\nabla_{\partial_a}\partial_b)^T}N\Bigr)$, $g(R(N,\partial_a)\partial_b,N)$ has been denoted by $Ric(N)$. $A(N)=h^{ab}K(\partial_a,W_{N}(\partial_b))$ is the Simon's operator whereas  $C^\perp$ is defined as $C^\perp=h^{ab}C(\partial_a,\partial_b)^\perp$ and $\tilde{H}=\accentset{(1)}{P}^{ab}K(\partial_{a},\partial_{b})$.
Thus identifying the Jacobi/stability operator $(\mathcal L)$ for minimal surfaces as
\begin{gather}
 \mathcal{L}N=\Delta^{\perp}N+Ric(N)+A(N),
\end{gather}
we can rewrite \eqref{bang} as
\begin{gather}\label{devi}
  \mathcal{L}N=-C^\perp+\tilde{H}.
\end{gather}
This is the inhomogeneous Jacobi equation. The solutions of this equation will provide us with the deformation of a minimal surface under a perturbation of the ambient spacetime. The inhomogeneous terms in the above equation, involves perturbation of the metric and is the only term in the above equation that involves the perturbation. If there were no perturbations the equation would have corresponded to the one describing  a deviation of a minimal surface to another minimal surface in the same spacetime $(\mathcal M_0,g_0)$. We will solve for solutions of this equation for specific cases and substitute the result in an area variation formula which we derive in the next section.
%\section{Solution of the inhomogeneous Jacobi equation}
%We will consider $AdS_4$ as our unperturbed background spacetime
\section{Variation of the Area functional}
According to Hubeny, Rangamani, Takayanagi (HRT) proposal the area of a codimension two spacelike extremal surface$(\gamma_A)$ in $AdS_{d+1}$ whose boundary coincides with the boundary of subsystem $A$ gives the entanglement entropy for this subsystem. Our goal therefore would be to obtain the change in area of a minimal surface up to second order with the extra constraint that the boundary of the surface remain unaltered i.e the deviations vanish at the boundary. At second order we will encounter terms which involve the deviation of the embedding functions itself. It is here that we have to use the solutions of the inhomogeneous Jacobi equation. The first variation of area of the minimal surface is given by,
\begin{gather}\label{c}
 \delta_NA=
 \int d^{n}\tau~{\frac{\sqrt{h}}{2}}h^{ab}\delta_N h_{ab}= -\int d^{n}\tau~\sqrt{h}g(N,H)+{1\over 2}\int d^{n}\tau~\sqrt{h} h^{ab}\accentset{(1)}{P}(\partial_{a},\partial_{b})+\text{Surface terms}.
\end{gather}
If the perturbations are set to zero then we get back the known expression for first variation of area. In the presence of perturbations the on-shell expression can be obtained by setting ($H=0$).
\begin{gather}
  \delta_NA={1\over 2}\int d^{n}\tau~\sqrt{h} h^{ab}\accentset{(1)}{P}(\partial_{a},\partial_{b})
\end{gather}
The second variation of area is given by
\begin{eqnarray}\label{svar}
 \delta^{(2)}_NA=-\int d^{n}\tau~\delta_N(\sqrt{h}g(N,H))+{1\over 2}\int d^{n}\tau~\delta_N(\sqrt{h} h^{ab}\accentset{(1)}{P}(\partial_{a},\partial_{b}))+\text{Surface terms}
\end{eqnarray}
Note that since $[N,\partial_a]=0$ for all $a$, the variation of the surface term is again a surface term. From the results of the previous section \ref{DOJE}, the first term in the above expression can be written in terms of the stability operator. Simplifying the second term requires a bit of algebra. Note that $\delta_N(\sqrt{h}h^{ab}\accentset{(1)}{P}(\partial_a,\partial_b))$ has the following expression,
%\begin{gather}\label{ver}
 %\delta_N(\sqrt{h}g(N,H))=\left(-\sqrt{h}g(N,H)+{1\over 2}\sqrt{h} h^{ab}P(\partial_{a},\partial_{b})\right)g(N,H)+\sqrt{h}P(N,H)\notag\\+\sqrt{h}g(\nabla_{N}N,H)+g(N,\delta_NH)
%\end{gather}
%and 
\begin{eqnarray}\label{ben}
 \sqrt{h}h^{ab}\accentset{(1)}{P}(\partial_a,\partial_b)\left(-g(N,H)+{1\over 2}h^{cd}\accentset{(1)}{P}(\partial_{c},\partial_{d})\right)+2\sqrt{h}h^{ac}h^{bd}g(N,K(\partial_c,\partial_d))\accentset{(1)}{P}(\partial_a,\partial_b)\nonumber\\
 -\sqrt{h}h^{ac}h^{bd}\accentset{(1)}{P}(\partial_c,\partial_d)\accentset{(1)}{P}(\partial_a,\partial_b)+\sqrt{h}h^{ab}\left[2\accentset{(1)}{P}(\nabla_{\partial_a}N,\partial_b)+2g(C(\partial_a,N),\partial_b)+\accentset{(2)}{P}(\partial_a,\partial_b)\right]
\end{eqnarray}

substituting the expression in (\ref{ben}) in (\ref{svar}) and using the conditions $H=0, \delta_NH=0$, one arrives at the following final expression for the second variation of the area functional \footnote{where we have used the following two expressions,
\begin{gather}
 (\nabla_{\partial_a}P)(N,\partial_b)=g(C(\partial_a,\partial_b),N)+g(C(\partial_a,N),\partial_b)\notag
\end{gather}
%Using this in the above expression we get
%\begin{gather}
 %\delta^{(2)}_NA={1\over 4}\int d^{n}x~\sqrt{h}h^{ab}P(\partial_a,\partial_b)h^{cd}P(\partial_{c},\partial_{d})\notag\\+\int d^{n}x~\sqrt{h}h^{ac}h^{bd}g(N,K(\partial_c,\partial_d))P(\partial_a,\partial_b)-{1\over 2}\int d^{n}x~\sqrt{h}h^{ac}h^{bd}P(\partial_c,\partial_d)P(\partial_a,\partial_b)\notag\\+\int d^{n}x~\sqrt{h}h^{ab}{1\over 2}P^{(2)}(\partial_a,\partial_b)+\int d^{n}x~\sqrt{h}h^{ab}\left[P(\nabla_{\partial_a}N,\partial_b)+ (\nabla_{\partial_a})P(N,\partial_b)-g(C(\partial_a,\partial_b),N)\right]\notag\\
%\end{gather}
%Using the minimality condition $H=0$.We get
%\begin{gather}
 %\delta^{(2)}_NA={1\over 4}\int d^{n}x~\sqrt{h}h^{ab}P(\partial_a,\partial_b)h^{cd}P(\partial_{c},\partial_{d})\notag\\+\int d^{n}x~\sqrt{h}h^{ac}h^{bd}g(N,K(\partial_c,\partial_d))P(\partial_a,\partial_b)-{1\over 2}\int d^{n}x~\sqrt{h}h^{ac}h^{bd}P(\partial_c,\partial_d)P(\partial_a,\partial_b)\notag\\+\int d^{n}x~\sqrt{h}h^{ab}{1\over 2}P^{(2)}(\partial_a,\partial_b)+\int d^{n}x~\sqrt{h}h^{ab}\left[\nabla_{\partial_a}(P(N,\partial_b))-P((\nabla_{\partial_a}\partial_b)^T,N)\right]\notag\\-\int d^{n}x~\sqrt{h}h^{ab}g(C(\partial_a,\partial_b),N)\notag\\
%\end{gather}
%Now we note that 
\begin{gather}
\nabla_{\partial_a}[\sqrt{h}h^{ab}P(N,\partial_a)]=\sqrt{h}h^{ab}\nabla_{\partial_a}[P(N,\partial_b)]-\sqrt{h}h^{ab}P(N,(\nabla_{\partial_a}\partial_b)^T)\notag
\end{gather}.}, 
%\begin{gather}
 %\delta^{(2)}_NA=-\int d^{n}x~\Biggl[\left(-\sqrt{h}g(N,H)+{1\over 2}\sqrt{h} h^{ab}P(\partial_{a},\partial_{b})\right)g(N,H)+\sqrt{h}P(N,H)\notag\\+\sqrt{h}g(\nabla_{N}N,H)+g(N,\delta_NH)\Biggr]+{1\over 2}\int d^{n}x~\Biggl[\sqrt{h}h^{ab}P(\partial_a,\partial_b)\left(-g(N,H)+{1\over 2}h^{cd}P(\partial_{c},\partial_{d})\right)\notag\\+2\sqrt{h}h^{ac}h^{bd}g(N,K(\partial_c,\partial_d))P(\partial_a,\partial_b)-\sqrt{h}h^{ac}h^{bd}P(\partial_c,\partial_d)P(\partial_a,\partial_b)\notag\\+\sqrt{h}h^{ab}\left[2P(\nabla_{\partial_a}N,\partial_b)+2g(C(\partial_a,N),\partial_b)+P^{(2)}(\partial_a,\partial_b)\right]\Biggr]
%\end{gather}
%Thus the onshell value can be obtained by setting()
%\begin{gather}
 %\delta^{(2)}_NA={1\over 4}\int d^{n}x~\sqrt{h}h^{ab}P(\partial_a,\partial_b)h^{cd}P(\partial_{c},\partial_{d})\notag\\+\int d^{n}x~\sqrt{h}h^{ac}h^{bd}g(N,K(\partial_c,\partial_d))P(\partial_a,\partial_b)-{1\over 2}\int d^{n}x~\sqrt{h}h^{ac}h^{bd}P(\partial_c,\partial_d)P(\partial_a,\partial_b)\notag\\+\int d^{n}x~\sqrt{h}h^{ab}\left[P(\nabla_{\partial_a}N,\partial_b)+g(C(\partial_a,N),\partial_b)+{1\over 2}P^{(2)}(\partial_a,\partial_b)\right]
%\end{gather}
%Now we can write
\begin{eqnarray}\label{svar2}
 \delta^{(2)}_NA&=&{1\over 4}\int d^{n}\tau~\sqrt{h}h^{ab}\accentset{(1)}{P}(\partial_a,\partial_b)h^{cd}\accentset{(1)}{P}(\partial_{c},\partial_{d})\notag\\
 &&+\int d^{n}\tau~\sqrt{h}h^{ac}h^{bd}g(N,K(\partial_c,\partial_d))\accentset{(1)}{P}(\partial_a,\partial_b)-{1\over 2}\int d^{n}\tau~\sqrt{h}h^{ac}h^{bd}\accentset{(1)}{P}(\partial_c,\partial_d)\accentset{(1)}{P}(\partial_a,\partial_b)\notag\\
 &&+\int d^{n}\tau~\sqrt{h}h^{ab}{1\over 2}\accentset{(2)}{P}(\partial_a,\partial_b)-\int d^{n}\tau~\sqrt{h}h^{ab}g(C(\partial_a,\partial_b),N)+\text{Surface terms},
\end{eqnarray}

The appearance of surface terms in the above expression is not very crucial, at least in the context of our current work. Since the boundary subsystem is kept fixed, while the bulk metric is being perturbed, the boundary conditions on the deviation vector would imply that it vanishes at the boundary. Thus change in area will have no contribution from the boundary terms. If we started with a more general deviation vector which also had components tangent to the immersed surface, then the only modification of the above expression would have been through the appearance of more boundary terms. The bulk contribution still would have arised from normal variations only. This will be shown in full rigor in  a later work \footnote{A.Ghosh and R.Mishra, work in progress}. where we will primarily use these boundary terms to find the change of entanglement entropy due to deformations of the subsystem itself.

\section{Brief outline of steps involved in obtaining Area variation upto second order}

 Our goal is to provide a formalism to calculate a change in the area of an extremal surface under changes of embedding and perturbation of metric. For the sake of brevity, all our calculations will be done in $3+1$ dimensions. But this can be easily generalized to higher dimensions. In this section, we provide a brief outline of this formalism
 
 1) Our first task is to take an asymptotically $AdS$ metric (to be considered as a perturbation over $AdS$) and identify the first and second order metric perturbations. In our case, this is achieved by writing the boosted AdS black brane metric in the Fefferman Graham coordinates, keeping up to second order (appendix \ref{Pert}). From the first order metric perturbations $\accentset{(1)}{P}_{\mu\nu}$ one can calculate the $(1,2)$ tensor.
 \begin{gather}
{C}^{\mu}_{\nu \rho}=\frac{1}{2}{g}^{\mu\sigma}\left(\partial_\nu \accentset{(1)}{P}_{\rho\sigma}+\partial_\rho \accentset{(1)}{P}_{\nu \sigma}-\partial_\sigma \accentset{(1)}{P}_{\nu \rho}\right)-\frac{1}{2}\accentset{(1)}{P}^{\mu \sigma}\left(\partial_\nu{g}_{\rho \sigma}+\partial_\rho{g}_{\nu \sigma}-\partial_\sigma{g}_{\nu \rho}\right),
\end{gather}
 where $g_{\mu\nu}$ is the unperturbed $AdS_4$ metric. The tensor defined is nothing but $C(X,Y)$ written in a coordinate system, i.e $C(\partial_\nu,\partial_\rho)=C^{\mu}_{\nu\rho}\partial_\mu$.
 
 2) Next we choose a free boundary extremal surface in $AdS_4$ \cite {Fonda:2014cca}. We will consider two cases A) half sphere in $AdS_4$ which is the corresponding minimal surface for a circular disc like subsystem and B) minimal surface corresponding to a thin strip boundary subsystem. With these choices and the choice of the perturbed metric $\accentset{(1)}{P}^{\mu\nu},$ we can now solve the inhomogeneous Jacobi equation \eqref{devi} and obtain the deviation vector ($N$).
 
 3) First and second order change in the area can be obtained by substituting the values of the deviation vector ($N$), first order metric perturbation $(\accentset{(1)}{P}_{\mu\nu})$  and the second order metric perturbation ($\accentset{(2)}{P}_{\mu\nu}, {C}^{\mu}_{\nu \rho}$) in the expression \eqref{svar2},\eqref{c} and integrating. From here the total change in area upto second order can be obtained as,
 \begin{gather}
  \Delta A =\Delta^{(1)}A+\frac{1}{2}\Delta^{(2)}A
  \end{gather}
 In the topic of the present paper we have selected asymptotically AdS spacetime. But this formalism can be easily applied to asymptotically flat case also. Here we have considered first order deviations of the extremal surface and second order metric perturbation to calculate the change in area up to second order. To calculate the change in area up to third order one need to consider second order deviation of the extremal surface and third order metric perturbations. Second order deviation can be obtained by extending the inhomogeneous Jacobi equation up to second order. The form of second order inhomogeneous Jacobi equation for geodesics can be found in \cite{Ghosh:2016fop}. Third order metric perturbation can be obtained by keeping third order terms in the asymptotic(Fefferman Graham) metric.

\section{Solutions of the inhomogeneous Jacobi equations and change in area}

 Our choice of the asymptotic metric to be considered as a perturbation over $AdS_4$ is the Boosted AdS black brane metric written in the Fefferman Graham coordinates upto second order. The CFT state dual to this bulk geometry is a thermal plasma which is uniformly boosted along a certain direction and is characterized by a temperature $T$ and boost $\beta$. This choice of a stationary spacetime is made to elucidate that our formalism can be easily applied to both static and non static spacetimes and yields expected results for the non-static case. The metric for $AdS_4$ in Poincar\'e coordinates reads as
 \begin{gather}
  ds^{2}= {-dt^2+dx^2+dy^2+dz^2\over z^2}
 \end{gather}
for simplicity we have set the radius of AdS to one. Now we will solve the inhomogeneous Jacobi equation and obtain an expression for the change in area for the case of two boundary subsystems namely
 
 \subsection{Circular disk subsystem}
In the case where the boundary subsystem is a circular disk of radius $\mathscr R$, it is known that the minimal surface in the $AdS_{d+1}$ is a $d-1$ dimensional hypersphere. The embedding of such a surface in $AdS_4$ is given by the following embedding functions \cite{Bhattacharya:2013bna, Fonda:2014cca},
 \begin{gather}\label{embedsphere}
 x=\mathscr R\sin{\theta}\cos{\phi}+X,~~y=\mathscr R\sin{\theta}\sin{\phi}+Y,~~z=\mathscr R\cos{\theta},~~t=constant.
 \end{gather}
The coordinates $\theta, \phi$ are the coordinates intrinsic to the surface and have ranges, $0\le\theta\le\frac{\pi}{2}$ and $0\le\phi<2\pi$. As is evident from the above expressions in eq.(\ref{embedsphere}), the surface of intersection of the half sphere with the $AdS_4$ boundary is at $\theta=\frac{\pi}{2}$. The intrinsic metric can be calculated via a pullback of the metric on the full space time and is given as, 
\begin{gather}
ds^2_{induced}=h_{ab}~dx^a dx^b={d\theta^2 +\sin^2{\theta} d\phi^2\over\cos^2{\theta}}
\end{gather}
To facilitate our calculation we will construct a local basis adapted to this surface. To start with we first construct a local tangent basis. As is apparent from the expression for the induced metric, the tangent bases are,
\begin{gather}
e_{2}=\cos{\theta}\partial_{\theta},~~e_{3}=\cot{\theta}\partial_{\phi}.
\end{gather}
Since the surface is purely spacelike, this set provides the space like bases for the full spacetime. The set of basis vectors spanning the normal bundle will provide us with the other two basis vectors. To obtain them we first lift the tangent vectors to the space time, by using the embedding functions and then use the orthogonality relations. As a matter of convention we mark the time like normal as $e_0$ and the space like normal as $e_1$.
\begin{gather}
e_{0}=z\partial_{t},~~e_{1}={z(x-X)\over l}\partial_{x}+{z(y-Y)\over l}\partial_{y}+{z^2\over l}\partial_{z}
\end{gather}
To completely specify the embedding one also needs to find the extrinsic curvatures and the intrinsic connection. To do so we need to find the covariant derivatives between the tangent vectors. They turn out to be,
\begin{gather}\label{ec}
\nabla_{e_2}e_2 =0,~~\nabla_{e_3}e_3 =-\cosec{\theta}~e_{2},~~\nabla_{e_3}e_2 =\cosec{\theta}~e_{3}
\end{gather}
which gives the following for the intrinsic connection and the extrinsic curvature.
\begin{gather}
D_{e_2}e_{2}=0,~~D_{e_3}e_{3}=-\cosec{\theta}~e_{2},~~D_{e_2}e_{3}=0,~~D_{e_2}e_{2}=\cosec{\theta}~e_{3},~~D_{e_2}e_{2}=\cosec{\theta}e_{3}\notag\\
K(e_2 ,e_2)=0,~~K(e_3 ,e_3)=0,~~K(e_2 ,e_3)=0,~~K(e_3 ,e_2)=0
\end{gather}
The vanishing of the extrinsic curvature implies that the surface is totally geodesic i.e any curve that is a geodesic on the surface is also a geodesic of the full spacetime. Recall that the Jacobi equation involves the connection in the Normal bundle $\nabla^\perp$, which can be found by calculating the covariant derivative of a normal vector along a tangent vector.
\begin{gather}\label{nb}
\nabla_{e_2}e_0=0,~~\nabla_{e_3}e_0=0,~~\nabla_{e_2}e_1=0,~~\nabla_{e_3}e_1=0
\end{gather}
From this one can read off the normal connection $\nabla^\perp$, using the Weingarten map. The procedure involves expanding the normal connection as $\nabla^\perp_{e^a}e_{A}=\beta_A^B(e^a)e_{B}$ ($A,~B$ denotes an index for basis vectors in the normal bundle) and yields,
\begin{gather}
\nabla^{\perp}_{e_2}e_0=\beta_{0}^{0}(e_2)e_0 +\beta_{0}^{1}(e_2)e_1 =0,~~\nabla^{\perp}_{e_3}e_0=\beta_{0}^{0}(e_3)e_0 +\beta_{0}^{1}(e_3)e_1 =0\notag\\
\nabla^{\perp}_{e_2}e_1=\beta_{1}^{0}(e_2)e_0 +\beta_{1}^{1}(e_2)e_1 =0,~~\nabla^{\perp}_{e_3}e_1=\beta_{1}^{0}(e_3)e_0 +\beta_{1}^{1}(e_3)e_1 =0.
\end{gather}
The vanishing of the $\beta's$ is equivalent to saying that the normal bundle is flat. Using the above results, calculating the left hand side of the Jacobi equation is just a matter of algebra. We expand the deviation vector in the normal basis as $\alpha^A~e_A$ and find the following equations for the $\alpha^A$.
\begin{gather}
\cos^2{\theta}\partial^{2}_{\theta}\alpha^A+\cos^{2}{\theta}\cot{\theta}\partial_{\theta}\alpha^A+\cot^{2}{\theta}\partial^{2}_{\phi}\alpha^A-2\alpha^A=F^A,
\end{gather}
Where $F^A$ has been defined for compactness of the above expression and is given as in $F^A= e^A_\mu\big( C^{\perp\mu}+\tilde H^{\mu}\big)$. Note that in this case the both the normal projections yield the one and the same equation. The source of this symmetry can be traced back as due to the symmetry of the embedding surface itself. Before proceeding to find solutions of the above equation, we need to analyze the homogeneous equations. In other words we will impose the boundary condition that the deviation vector is zero at the boundary and check if this implies that the only solution of the `homogeneous' piece of the above equation is the trivial solution. As we will see, this knowledge would be helpful in our effort to obtain solutions of the `inhomogeneous' equations. The homogeneous equation can be solved by the method of separation of variables $\alpha^A(\theta,\phi)=\Theta^A(\theta)~\Phi^A(\phi)$. The equations then become ordinary differential equations.
\begin{gather}
{d^2\Theta^A\over d\theta^2}+\cot \theta{d\Theta^A\over d\theta}-(2\sec^2 \theta+m^2 \cosec^2~\theta)\Theta^A=0
\end{gather}
and the $\phi$ equation is,
\begin{gather}
{d^{2}\Phi^A\over d\phi^2}+m^2\Phi^A=0
\end{gather}
For the $\phi$ equation the boundary condition is of course the periodic one $\Phi^A(\phi+2\pi)=\Phi^A(\phi)$, which restricts the values of $m$ to integers only. The most general solution of this equation is given by,
\begin{gather}
\Theta=C_1~\cos^2{\theta}(\sin{\theta})^m~
_2F_1\Big(1+\frac{m}{2},\frac{3}{2}+\frac{m}{2};m+1;\sin^2{\theta}\Big)\notag\\
~~~~~~~~~~~~~~~~~~~~~~~~~~~~~~~~~~~~~~~+C_2~\cos^2{\theta}(\sin{\theta})^{-m}~
_2F_1\Big(1-\frac{m}{2},\frac{3}{2}-\frac{m}{2};-m+1;\sin^2{\theta}\Big)
\end{gather}
Assuming the boundary condition $\Theta=0$ at $\theta=\frac{\pi}{2}$ and demanding that the solution be regular at $\theta=0$, one concludes that $C_1=C_2=0$. To check this assume $m$ to be positive (Similar arguments would hold for $m$ negative). Note that at $\theta=0$ the second solution diverges since $_2F_1\Big(1-\frac{m}{2},\frac{3}{2}-\frac{m}{2};-m+1;0\Big)=1$, while the $\sin^{-m}(\theta)$ term diverges. This implies $C_2$ must be set to zero. At $\theta=\frac{\pi}{2}$ the first solution diverges. This can be argued in the following way. Note that $lim_{z\rightarrow 1^-}\frac{_2F_1\Big(a,b;c;z\Big)}{(1-z)^{c-a-b}}=\frac{\Gamma(c)\Gamma(a+b-c)}{\Gamma(a)\Gamma(b)}$ for $\Re(c-a-b)<0$. Writing the first solution as,
\begin{eqnarray}
\frac{z^{\frac{m}{2}}}{(1-z)^{\frac{1}{2}}}\frac{_2F_1\Big(1+\frac{m}{2},\frac{3}{2}+\frac{m}{2};m+1;z\Big)}{(1-z)^{-\frac{3}{2}}},
\end{eqnarray}
one can realize that solution is divergent at $\theta=\frac{\pi}{2}$. Hence $C_1$ has to be set to zero. As expected for homogeneous spaces the only solution is the trivial one.
%For the $m=0$ case the solutions are,
%\begin{gather}
%\Theta=\frac {C_1}{\cos{\theta} }+\frac {C_2 \left( -\cos{\theta} + \tanh^{-1} \left( \frac{1}{\cos{\theta}}  \right)\right) }{\cos \left( \theta \right) }
%\end{gather}

Now we will solve the inhomogeneous equation. By substituting $C\equiv\left(\frac{1}{3}+\beta^2\gamma^2\right)\frac{1}{z_0^3}$, $D\equiv\left(\frac{1}{3}\right)\frac{1}{z_0^3}$, $B\equiv\beta\gamma^2\frac{1}{z_0^3}$, and writing $\mathcal R^3\equiv\frac{\mathscr R^3}{z_0^3}$, the inhomogeneous equation for $e_1$ turns out to be,
\begin{gather}
\cos^2{\theta}\partial^{2}_{\theta}\alpha^1+\cos^{2}{\theta}\cot{\theta}\partial_{\theta}\alpha^1+\cot^{2}{\theta}\partial^{2}_{\phi}\alpha^1-2\alpha^1=\mathcal R^3\cos^4\theta\bigg(\frac{2}{3}+\beta^2\gamma^2\bigg)+\frac{5\mathcal R^3\sin^2\theta\cos^4\theta}{6}\notag\\
+\frac{5\mathcal R^3\beta^2\gamma^2\sin^2\theta\cos^4\theta}{4}+\frac{5\mathcal R^3\beta^2\gamma^2\sin^2\theta\cos^4\theta\cos2\phi}{4},
\end{gather}
and that for $e_0$ reads,
\begin{gather}
\cos^2{\theta}\partial^{2}_{\theta}\alpha^0+\cos^{2}{\theta}\cot{\theta}\partial_{\theta}\alpha^0+\cot^{2}{\theta}\partial^{2}_{\phi}\alpha^0-2\alpha^0=3\beta\gamma^2\mathcal R^3\cos^4\theta\sin\theta\cos\phi
\end{gather}
Let us consider the $e_1$ equations first. Note that since the equation is linear one can find the solutions for individual terms in the inhomogeneous piece separately. Let us therefore consider the terms containing no function of $\phi$.
\begin{gather}
\partial^{2}_{\theta}\alpha^1+\cot{\theta}\partial_{\theta}\alpha^1+\cosec^{2}{\theta}\partial^{2}_{\phi}\alpha^1-2~\sec^2\theta~\alpha^1=\mathcal R^3\cos^4\theta\bigg(\frac{2}{3}+\beta^2\gamma^2\bigg)+\frac{5\mathcal R^3\sin^2\theta\cos^4\theta}{6}\notag\\
+\frac{5\mathcal R^3\sin^2\theta\cos^4\theta\beta^2\gamma^2}{4}%+\frac{5\mathcal R^3\sin^2\theta\cos^4\theta\beta^2\gamma^2\cos2\phi}{4},
\end{gather}
Owing to the fact that the  right hand side of this equation contains no function of $\phi$ the only non trivial solution to this equation will come from $m=0$. This can be understood by taking a trial solution of the form $\sum_{m}\big(g_m(\theta)e^{im\phi}+g_{-m}(\theta)e^{-im\phi}\big)$. If one now lists the equations for individual $m's$, then only the $m=0$ equation will have an inhomogeneous term on the right hand side, while the other equations will be all homogeneous. But we have already shown that the solutions of the homogeneous equations are trivial. Therefore we only need to solve the $m=0$ equation, which reads,
\begin{gather}
\frac{d^{2}\Theta^1}{d{\theta}^2}+\cot{\theta}\frac{d\Theta^1}{d\theta}-2~\sec^2\theta~\Theta^1=\mathcal R^3\cos^2\theta\bigg(\frac{2}{3}+\beta^2\gamma^2\bigg)+\frac{5\mathcal R^3\sin^2\theta\cos^2\theta}{6}
+\frac{5\mathcal R^3\sin^2\theta\cos^4\theta\beta^2\gamma^2}{4}%+\frac{5\mathcal R^3\sin^2\theta\cos^4\theta\beta^2\gamma^2\cos2\phi}{4},
\end{gather}
The solution to this equation with the conditions that it is zero at $\theta=\frac{\pi}{2}$ and regular at $\theta=0$ is given by,
\begin{gather}
\Theta^1=\frac{1}{288} \mathcal R^3 \cos ^2\theta\bigg(3 \beta ^2 \gamma ^2+2\bigg)  \bigg(3 \cos 2\theta-23\bigg)
\end{gather}
The other equation containing a $\cos2\phi$ is equivalent to solving the $\theta$ equation for $m=2$.
\begin{gather}
\partial^{2}_{\theta}\Theta^1+\cot{\theta}\partial_{\theta}\Theta^1-4~\cosec^{2}{\theta}~\Theta^1-2~\sec^2\theta~\Theta^1=\frac{5\mathcal R^3\beta^2\gamma^2\sin^2\theta\cos^2\theta}{4},
\end{gather}
The solution to this equation with conditions as above yields,
\begin{gather}
\Theta^1=-\frac{1}{64} \mathcal R^3 \beta^2 \gamma^2 \bigg(\sin{2\theta}\bigg)^2
\end{gather}
The full solution is then,
\begin{gather}
\alpha^1=\frac{1}{288} \mathcal R^3 \cos ^2\theta\bigg(3 \beta ^2 \gamma ^2+2\bigg)  \bigg(3 \cos 2\theta-23\bigg)-\frac{1}{64} \mathcal R^3 \beta^2 \gamma^2 \bigg(\sin{2\theta}\bigg)^2\cos2\phi
\end{gather}
Now, we go over to the $e_0$ equation. By similar arguments, one concludes that the only contribution to the solution will come from $m=1$ term. Therefore, the equation becomes, 
\begin{gather}
\partial^{2}_{\theta}\alpha^0+\cot{\theta}\partial_{\theta}\alpha^0-\cosec^{2}{\theta}~\alpha^0-2~\sec^2\theta~\alpha^0=3\beta\gamma^2\mathcal R^3\cos^2\theta\sin\theta
\end{gather}
Along with the usual boundary conditions, the solution to this equation is,
\begin{gather}
\alpha^0=-\frac{1}{4} \beta \gamma ^2 \mathcal R^3 \sin \theta \cos ^2\theta\cos{\phi}
\end{gather}
%%%%%%%%%%%%%%%%%%%%%%%%%%%%%%%%%%%%%%%%%%%%%%%%%%%%%%%%%%%%%%%%%%%%%%%%%%%%%%%%%%%%%%%%%%%%%%%%%%%%%%%%%%%%%%%%%%%%%%%%%%%%%%%%%%%%%%%%%%%%%%%%%%%%%%%%%%%%%%%%%%%%%%%%%%%%%%%%%%%%%%%%%%%%%%%%%%%%%%%%%%%%%%%%%%%%%%%%%%%%%%%%%%%%%%%%%%%%%%%%%%%%%%%%%%%%%%%%%%%%%%%%%%%%%%%%%%%%%%%%%%%%%%%%%%%%%%%%%%%%%%%%
The very fact that the solution of the above $e_0$ equation is non trivial proves the fact that the perturbed minimal surface ceases to be on a constant $t$ slice as was initially the case with the unperturbed minimal surface in $AdS_4$ background. One can also check that setting $\beta=0$, which gives the static case of an AdS Black Brane, makes $\alpha^0$ vanish.

We are now in a position to calculate the change in area. We first calculate the first order change in the area. As is known, at this order there is no contribution from deviations of the minimal surface itself, and therefore at this order the change must match with that obtained in \cite{Blanco:2013joa}. The first order change in HEE(S) for the spherical entangling surface can be extracted from eq.(\ref{c}) and is given by,

\begin{gather}
\Delta^{(1)}S={1\over 4G_N}\Delta^{(1)}A={1\over 8G_N} \int d^{d-1}\tau~\sqrt{h}h^{ab}\accentset{(1)}{P}(\partial_{a},\partial_{b})%\notag\\
%=\int_{0}^{\frac{\pi}{2}}\int_{0}^{2\pi}\frac{1}{24} L^3 \cos ^3\theta \bigg(\big(3 \beta ^2 \gamma ^2+2\big) \big(\cos 2\theta+3\big)-6 \beta ^2 \gamma ^2 \sin ^2\theta \cos 2\theta\bigg)~\frac{\sin \theta}{\cos ^2\theta}~d\theta~d\phi\notag\\
=\frac{1}{32G_N} \pi \mathscr R^3\left(3 \beta ^2 \gamma ^2+2\right)\frac{1}{z_0^3}
 \end{gather}
 The second order variation has contributions from various terms. The full expression is given by eqn.(\ref{svar2}),
 \begin{gather}\label{areachange2ndorder}
 \Delta^{(2)}A=\int d^{d-1}\tau~\sqrt{h} \Bigl(h^{ab}h^{cd}\accentset{(1)}{P}(\partial_b,\partial_d)g(N^{\perp},K(\partial_a,\partial_c))-h^{ab}g(C(\partial_a,\partial_b),N^{\perp})\Bigr)\notag\\
 +\int d^{d-1}\tau~\sqrt{h}\Bigl[{h^{ab}\over 2}\accentset{(2)}{P}(\partial_{a},\partial_{b})-{1\over 2}h^{ac}h^{bd}\accentset{(1)}{P}(\partial_{a},\partial_{b})\accentset{(1)}{P}(\partial_{c},\partial_{d})+{1\over 4}h^{ab}h^{cd}\accentset{(1)}{P}(\partial_{c},\partial_{d})\accentset{(1)}{P}(\partial_{a},\partial_{b})\Bigr]
 \end{gather}
 Let us analyze the above equation. The last three terms in the above equation eq.(49) are the terms coming purely from the bulk metric perturbations. The first and the second term arise from changes due change in the embedding function itself. The $N^\perp$ in the above equation therefore has to be substituted with the solutions of the Jacobi equation obtained before and then the integrals calculated. We therefore enumerate the results one by one. Consider the last three terms in the above expression which do not involve the deviation vector.
 
\begin{gather}
\int d^{d-1}\tau~\sqrt{h}{h^{ab}\over 2}\accentset{(2)}{P}(\partial_{a},\partial_{b})%\notag\\
%=\int_{0}^{\frac{\pi}{2}}\int_{0}^{2\pi}\frac{1}{288} \mathcal R^6 \cos ^6\theta \bigg\{12 \beta ^2 \gamma ^2 \sin ^2\theta \cos 2\phi-\big(6 \beta ^2 \gamma ^2-1\big) \big(\cos 2\theta +3\big)\bigg\}~\frac{\sin \theta}{\cos ^2\theta}~d\theta~d\phi\notag\\
=-\frac{1}{105} \pi  \mathcal R^6 \left(6 \beta ^2 \gamma ^2-1\right)
\end{gather}
The next term is a product of two metric perturbations gives,
 \begin{gather}
 \int d^{d-1}\tau~\sqrt{h}{1\over 2}h^{ac}h^{bd}\accentset{(1)}{P}(\partial_{a},\partial_{b})\accentset{(1)}{P}(\partial_{c},\partial_{d})%\notag\\
 %=\int_{0}^{\frac{\pi}{2}}\int_{0}^{2\pi}\frac{1}{18} \mathcal R^6 \cos ^6\theta \Bigg[\cos ^4\theta \bigg\{\big(3 \beta ^2 \gamma ^2+1\big)^2 \cos ^4\theta+1\bigg\}+\sin ^4\theta \bigg\{\big(3 \beta ^2 \gamma ^2+1\big)^2+\cos ^4\theta\bigg\}\notag\\
 %+2 \sin ^2\theta \cos ^2\theta \bigg\{3 \beta ^2 \gamma ^2+9 \beta ^4 \gamma ^4 \cos ^2\theta+\big(3 \beta ^2 \gamma ^2+1\big) \cos ^4\theta+1\bigg\}\Bigg]~\frac{\sin \theta}{\cos ^2\theta}~d\theta~d\phi\notag\\
 =\frac{2 \pi  \mathcal R^6 \left(216 \beta ^4 \gamma ^4+147 \beta ^2 \gamma ^2+49\right)}{2835}
 \end{gather}
Finally the other term containing a product of two perturbations evaluates to,
 \begin{gather}
 \int d^{d-1}\tau~\sqrt{h}{1\over 4}h^{ab}h^{cd}\accentset{(1)}{P}(\partial_{c},\partial_{d})\accentset{(1)}{P}(\partial_{a},\partial_{b})=\frac{2 \pi  \mathcal R^6 \left(108 \beta ^4 \gamma ^4+141 \beta ^2 \gamma ^2+47\right)}{2835}
 \end{gather}
Note that the contribution from the first term is zero owing to the fact that the extrinsic curvature $K(\partial_a,\partial_b)$ is zero in this case of a spherical boundary subsystem. As we will see later this term does give non zero contributions for the case of a strip subsystem. While calculating the second term, the $N^\perp$ contained in the term has to be substituted with the solutions of the inhomogeneous stability equation. After substitution one obtains,
 \begin{gather}
 \int d^{d-1}\tau~\sqrt{h} h^{ab}g(C(\partial_a,\partial_b),N^{\perp})%\notag\\
= \frac{\pi  \mathcal R^6 \left(459 \beta ^4 \gamma ^4+\beta ^2 \left(81 \gamma ^4+597 \gamma ^2\right)+199\right)}{1890}
 \end{gather}
 The total second order Change in HEE is then given by,
 \begin{gather}\label{sed}
\Delta^{(2)}S={1\over 4G_N} \Delta^{(2)}A=-{\pi \mathscr R^6\over 4G_N}\frac{\left(1809 \beta ^4 \gamma ^4+3 \beta ^2 \left(81 \gamma ^2+713\right) \gamma ^2+551\right)}{5670}\frac{1}{z_0^6}
 \end{gather}
 This expression gives the second order change of HEE. Positivity of relative entropy between two states in the CFT demands that 
 \begin{gather*}
  \Delta H\geq\Delta S
 \end{gather*}
 Where $H$ is the modular Hamiltonian for the spherical entangling surface, given in terms of the boundary stress tensor. One can now check that the equality is satisfied at the first order \cite{Blanco:2013joa}. As the modular Hamiltonian remains unchanged at second order, positivity of relative entropy demands that $\Delta^{(2)}S\leq 0$ at second order. Our result eq (\ref{sed}) is therefore in agreement with this observation.
The full expression for change of HEE is then given by 
\begin{gather}
 \Delta S=\Delta^{(1)}S+\frac{1}{2}\Delta^{(2)}S\notag\\
 =\frac{1}{32G_N} \pi \mathscr R^3\left(3 \beta ^2 \gamma ^2+2\right)\frac{1}{z_0^3}-{\pi \mathscr R^6\over 8G_N}\frac{\left(1809 \beta ^4 \gamma ^4+3 \beta ^2 \left(81 \gamma ^2+713\right) \gamma ^2+551\right)}{5670}\frac{1}{z_0^6}
\end{gather}
the above expression gives the net change in HEE for spherical entangling surface upto second order over pure $AdS$(ground state) value.
 %%%%%%%%%%%%%%%%%%%%%%%%%%%%%%%%%%%%%%%%%%%%%%%%%%%%%%%%%%%%%%%%%%%%%%%%%%%%%%%%%%%%%%%%%%%%%%%%%%%%%%%%%%%%%%%%%%%%%%%%%%%%%%%%%%%%%%%%%%%%%%%%%%%%%%%%%%%%%%%%%%%%%%%%%%%%%%%%%%%%%%%%%%%%%%%%%%%%%%%%%%%%%%%%%%%%%%%%%%%%%%%%%%%%%%%%%%%%%%%%%%%%%%%%%%%%%%%%%%%%%%%%%%%%%%%%%%%%%%%%%%%%%%%%%%%%%%%%%%%%%%%%%%%%%%%%%%%%%%%%%%%%%%%%%%%%%%%%%%%%%%%%%%%%%%%%%%%%%%%%%%%%%%%%%%%%%%%%%%%%%%%%%%%%%%%%%%%%%%%%%%%%%%%%%%%%%%%%%%%%%%%%%%%%%%%%%%%%%%%%%%%%%%%%%%%%%
 \subsection{Thin Strip subsystem}
 We now consider a two dimensional strip like subsystem on the $AdS_4$ boundary. The subsystem is given by the region $[-L,L]\times[-\frac{l}{2},\frac{l}{2}]$ of the $x-y$ plane, where $L\gg l$. The minimal surface corresponding to such a subsystem  \cite {Fonda:2014cca} is characterized by the following embedding functions,
 \begin{gather}
 x=\lambda,~~y(\theta)=-z_* E\left({(\pi-2\theta)\over 4}\left.\right|2\right),~~z(\theta)=z_*\sqrt{\sin{\theta}},
\end{gather}
where $z_{*}$ is the turning point of the minimal surface in $AdS_4$ and $E(\alpha,\beta)$ is the incomplete elliptic integral of the second kind. Note that due to the condition $L\gg l$ the effects of the sides of the minimal surface can be neglected. The embedding function clearly reflects this approximation. In intrinsic coordinates the metric takes the form
\begin{gather}
{ds^2}_{induced}={{z_*}^2d\theta^2+4\sin{\theta}d\lambda^2\over 4 z_*^2\sin{\theta}^2},
\end{gather}
the range of the coordinates being $0\le\theta\le\pi$ and $-L\le\lambda\le L$. Further the turning point $z_{*}$ can be written in terms of the width $l$ of the subsystem as $z_*=\frac{\Gamma(\frac{1}{4})l}{2\sqrt{\pi}\Gamma(\frac{3}{4})}$. We also need to calculate the extrinsic curvature and the connection in the normal bundle. We again use a local tetrad adapted to the surface. The two spacelike bases are chosen such that they are tangent to the embedded surface. In intrinsic coordinate, they have the form,
\begin{gather}
 e_2=2\sin{\theta}\partial_{\theta},~~e_3=z_*\sqrt{\sin{\theta}}\partial_{\lambda}
\end{gather}
These are lifted to the full spacetime coordinates and then by using orthogonality relations one can construct the bases which span the normal bundle.
\begin{gather}
 e_1=z(\sin{\theta}\partial_z-\cos{\theta}\partial_y),~~e_0=z\partial_t
\end{gather}
The covariant derivatives of the normal vectors are given by,
\begin{gather}
 \nabla_{e_2}e_1=\sin{\theta}~e_2,~~\nabla_{e_3}e_1=-\sin{\theta}~e_3,~~\nabla_{e_2}e_0=0,~~\nabla_{e_3}e_0=0
\end{gather}
From these one can read of the Weingarten maps and therefore the extrinsic curvatures,
\begin{gather}
 W_{e_1}(e_2)=-\sin{\theta}~e_2,~~W_{e_1}(e_3)=\sin{\theta}~e_3,~~W_{e_0}(e_2)=0,~~W_{e_0}(e_3)=0
\end{gather}
We are now in a position to calculate the left hand hand side of the Jacobi equation. We expand the deviation vector as $\alpha^A~e_A$ and then by using the above expressions we get,
\begin{gather}
 4\sin^2{\theta}~\partial^2_\theta\alpha^1+2\sin{\theta}\cos{\theta}~\partial_{\theta}\alpha^1+{z_*}^2\sin{\theta}~\partial^2_\lambda\alpha^1-2\cos^2{\theta}~\alpha^1=F^1\notag\\
4\sin^2{\theta}~\partial^2_\theta\alpha^0+2\sin{\theta}\cos{\theta}~\partial_{\theta}\alpha^0+{z_*}^2\sin{\theta}~\partial^2_\lambda\alpha^0-2\alpha^0=F^0
\end{gather}
As before, we first analyze the homogeneous equations by solving them using separation of variables.
\begin{gather}
{d^2\Theta^{1}\over d\theta^2}+\frac{1}{2}\cot{\theta}{d\Theta^{1}\over d\theta}-\bigg(\frac{1}{2}\cot^2{\theta}+\frac{k^2}{4\sin\theta}\bigg)\Theta^{1}=0\notag\\
{d^2\Theta^{0}\over d\theta^2}+\frac{1}{2}\cot{\theta}{d\Theta^{0}\over d\theta}-\bigg(\frac{1}{2}\cosec^2{\theta}+\frac{k^2}{4\sin\theta}\bigg)\Theta^{0}=0\notag\\
{d^{2}\Phi^{(0,1)}\over d\lambda^2}+\bigg({k\over z_*}\bigg)^2\Phi^{(0,1)}=0
\end{gather}
The solution to the $\theta$ part is given in terms of the generalized Heun's function,
%\begin{gather}
%\Theta^1=\frac{C_1\bigg(1+\sin{\theta}\bigg)^{\frac{1}{4}}\bigg(1-\sin{\theta}\bigg)^{\frac{3}{4}}H\ell \bigg( -1,\frac{5}{4}+\frac{1}{4}{m}^{2};1,\frac{5}{2},\frac{5}{2},\frac{1}{2}
%;-\sin\theta\bigg)}{\sqrt{\cos\theta}}\notag\\
%~~~~~~~~~~~~~~~~~~~~~~~~~~~~~~~~+\frac{C_2\bigg(1+\sin{\theta}\bigg)^{\frac{1}{4}}\bigg(1-\sin{\theta}\bigg)^{\frac{3}{4}}H\ell \bigg( -1,-\frac{1}{4}+\frac{1}{4}{m}^{2};-\frac{1}{2},1,-\frac{1}{2},\frac{1}{2}
%;-\sin\theta\bigg)}{\sqrt{\cos\theta\sin\theta}}\\
%\Theta^0=\frac{C_1\bigg(1+\sin{\theta}\bigg)^{\frac{1}{4}}\bigg(1-\sin{\theta}\bigg)^{\frac{3}{4}}H\ell \bigg( -1,\frac{5}{4}+\frac{1}{4}{m}^{2};\frac{3}{2},2,\frac{5}{2},\frac{1}{2}
%;-\sin\theta\bigg)}{\sqrt{\cos\theta}}\notag\\
%~~~~~~~~~~~~~~~~~~~~~~~~~~~~~~~~+\frac{C_2\bigg(1+\sin{\theta}\bigg)^{\frac{1}{4}}\bigg(1-\sin{\theta}\bigg)^{\frac{3}{4}}H\ell \bigg( -1,-\frac{1}{4}+\frac{1}{4}{m}^{2};0,\frac{1}{2},-\frac{1}{2},\frac{1}{2}
%;-\sin\theta\bigg)}{\sqrt{\cos\theta\sin\theta}}
%\end{gather}
%%%%%%%%%%%%%%%%%%%%%%%%%%%%%%%%%%%%%%%%%%%%%%%%%%%%%%%%%%%%%%%%%%%%%%%%%%%%%%%%%%%%%%%%%%%%%%%%%%%%%%%%%%%%%%%%%%%%%%%%%%%%%%%%%%%%%%%%%%%%%%%%%%%%%%%%%%%%%%%%%%%%%%%%%%%%%%%%%%%%%%%%%%%%%%%%%%%%%%%%%%%%%%%%%%%%%%%%%%%%%%%%%%%%%%%%%%%%%%%%%%%%%%%%%%%%%%%%%%%%%%%%%%%%%%%%%%%%%%%%%%%%%%%%%%%%%%%%%%%%%%%%%%%%%%%%%%%%%%%%%%%%%%%%%%%%%%%%%%%%%%%%%%%%%%%%%%%%%%%%%%%%%%%%%%%%%%%%%%%%%%%%%%%%%%%%%%%%%%%%%%%%%%%%%%%%%%%%%%%%%%%%%%%%%%%%%%%%%%%%%%%%%%%%%%%%%%%%%%%%%
and can be shown to yield trivial solutions under the boundary conditions assumed. We will now solve the inhomogeneous Jacobi equation for the strip subsystem for two separate cases ,

1. Strip along $'x'$ boost along $'x'$: In this case we consider the width of the strip to be along the $y$ direction and length along the $x$ direction in bounday of $AdS_4$. The inhomogeneous term for the Jacobi equation in this case is calculated for the asymptotic Boosted AdS blackbrane geometry  (appendix \ref{Pert}) where the boost is along the $x$ direction. 

2. Strip along $'x'$ boost along $'y'$: In this case the direction of the strip remains unchanged but the inhomogeneous term is now calculated for the same geometry but with the boost being along $y$ direction.

Changing the boost direction results in different deformations of the minimal surface. In the first case the surface remains on the same constant time ($t$) slice while in the second case there is a deviation of the surface along the time direction.

\subsubsection{\underline{Strip along `x' boost along `x'}}\label{SAXBAX}
In this case the $e_0$ equation turns out to be trivial i,e the inhomogeneous term is zero in the $e_0$ equation. Hence the surface remains on the same time slice. The $e_1$ equation is however non trivial. Note that since the right hand side is not a function of $\lambda$, only the  $k=0$ solution will be non trivial, which can be recast into,
\begin{gather}
{d^2\Theta^{1}\over d\theta^2}+\frac{1}{2}\cot{\theta}{d\Theta^{1}\over d\theta}-\bigg(\frac{1}{2}\cot^2{\theta}\bigg)\Theta^{1}=\frac{1}{4}\left(3D+\frac{3C}{2}\right)z_*^3\left(\sin{\theta}\right)^{\frac{1}{2}}-\frac{7}{8}Dz_*^3\left(\sin{\theta}\right)^{\frac{5}{2}},
\end{gather}
where expressions for $C,D$ can be found in appendix \ref{Pert}. The homogeneous solutions for this is,
\begin{gather}
\Theta^1(\theta) %= \frac{C_1\sin2\theta}{(2-2\cos2\theta)^{\frac{3}{4}}}+C_2\sqrt{1-\cos2\theta}~_2F_1\bigg(\frac{1}{4}, 1; %\frac{1}{2}, \frac{1}{2}\cos2\theta+\frac{1}{2}\bigg)\notag\\
= \frac{C_1\cos\theta}{\sqrt{\sin\theta}}+C_2\sin{\theta}~~_2F_1\bigg(\frac{1}{4}, 1; \frac{1}{2}, \cos^2\theta\bigg),
\end{gather}
and the Wronskian is $W(\theta)=e^{-\frac{1}{2}\int\cot(\theta)d\theta}=\frac{1}{\sqrt{\sin{\theta}}}$.
%\begin{gather}
%W(\theta)=e^{-\frac{1}{2}\int\cot(\theta)d\theta}=\frac{1}{\sqrt{\sin{\theta}}}
%\end{gather}
The full solution is then $\Theta^1_c+\Theta^1_p$.
\begin{gather}
\Theta^1=\frac{C_1\cos(\theta)}{\sqrt{\sin(\theta)}}+C_2\sin{\theta}~~_2F_1\bigg(\frac{1}{4}, 1; \frac{1}{2}, \cos^2\theta\bigg)\notag\\
-\frac{\cos(\theta)}{\sqrt{\sin(\theta)}}\int_{}^{\theta}\left[\frac{1}{4}\left(3D+\frac{3C}{2}\right)z_*^3\left(\sin{\theta}\right)^{\frac{1}{2}}-\frac{7}{8}Dz_*^3\left(\sin{\theta}\right)^{\frac{5}{2}}\right](\sin{\theta'})^{\frac{3}{2}}~~_2F_1\bigg(\frac{1}{4}, 1; \frac{1}{2}, \cos^2\theta'\bigg)d\theta'\notag\\
+\sin{\theta}~~_2F_1\bigg(\frac{1}{4}, 1; \frac{1}{2}, \cos^2\theta\bigg)\int_{}^{\theta}\left[\frac{1}{4}\left(3D+\frac{3C}{2}\right)z_*^3\left(\sin{\theta}\right)^{\frac{1}{2}}-\frac{7}{8}Dz_*^3\left(\sin{\theta}\right)^{\frac{5}{2}}\right]\cos(\theta')d\theta'
\end{gather}
It is not possible to get an analytical form of the integral involving the hypergeometric function. However since certain definite integrals are known for hypergeometric function, we hope that the final integral involving the change in area can be obtained by doing an integration by parts. %Note that the both the solutions are ill defined at the end points. Hence it would become immensely difficult to perform any mathematical procedures in this set up. To achieve what we set out for we therefore break the interval into $[0,\frac{\pi}{2}]$ and $[\frac{\pi}{2},\pi]$ calculating the area integrals for each half. Apart from the boundary conditions at $\theta=0,\pi$ we will assume that the solutions for each branch match at $\theta=\frac{\pi}{2}$. Consider first the $0\le\theta\le\frac{\pi}{2}$. The solution in this region can be written as,
%\begin{gather}
%\Theta^1=\frac{C_1\cos(\theta)}{\sqrt{2\sin(\theta)}}+C_2\sqrt{2}\sin{\theta}~~_2F_1\bigg(\frac{1}{4}, 1; \frac{1}{2}, \cos^2\theta\bigg)\notag\\
%-\frac{\cos(\theta)}{\sqrt{2\sin(\theta)}}\int_{0}^{\theta}f(\theta')~d\theta'+\sqrt{2}\sin{\theta}~~_2F_1\bigg(\frac{1}{4}, 1; \frac{1}{2}, \cos^2\theta\bigg)\frac{z_*^3 \sin^{\frac{3}{2}} \theta\big(-14 C+9 D \cos2\theta +75 D\big)}{42 \sqrt{2}}
%\end{gather}
To evaluate the integration constants we put the boundary condition $\Theta=0$ at $\theta=0$ and $\theta=\pi$. On demanding these the values of the constants turn out to be  $C_1=\frac{\pi z_*^3}{16} \left(2 C+D\right)$ and $C_2=-\frac{\Gamma \left(\frac{1}{4}\right)^2 z_*^3\left(2C+D\right)}{16 \sqrt{2 \pi }}$.
%that $\Theta=0$ at $\theta=0$ gives the following condition on $C_1$ and $C_2$.
%\begin{gather}
%C_1+\frac{2 \sqrt{2} \pi ^{3/2} C_2}{\Gamma \left(\frac{1}{4}\right)^2}=0
%\end{gather}
%One can evaluate the integral $\int_{0}^{\pi}f(\theta)d\theta$, the value turns out to be $-\frac{2\pi  z_*^3 (14 C-57 D)}{168 \sqrt{2}}$. Therefore the condition that $\Theta=0$ at $\theta=\pi$ implies the condition.
%\begin{gather}
%C_1-\frac{2 \sqrt{2} \pi ^{3/2} C_2}{\Gamma \left(\frac{1}{4}\right)^2}+\frac{2 \pi  z_*^3 (14 C-57 D)}{42 \sqrt{2}}=0,
%\end{gather}
%which further implies that $C_1=\frac{\sqrt{2} \pi  z_*^3}{84} \left(57  D-14  C\right)$ and $C_2=\frac{z_*^3 \Gamma \left(\frac{1}{4}\right)^2 (14 C-57 D)}{168 \sqrt{\pi }}$. Note that the third term in the above expression comes from the integral $\int_{0}^{\pi}f(\theta')d\theta'$.
%\begin{gather}
%\Theta=\frac{\cos(\theta)}{\sqrt{2\sin(\theta)}}\int_{0}^{\theta}f(\theta)~d\theta+\sqrt{2}\sin{\theta}~~_2F_1\bigg(\frac{1}{4}, 1; \frac{1}{2}, \cos^2\theta\bigg)\frac{z_*^3 \sin^{\frac{3}{2}} \theta\big(-14 C+9 D \cos2\theta +75 D\big)}{42 \sqrt{2}}
%\end{gather}

We now go over to the calculation of the integrals for calculating the change of area. Before calculating the terms involving the deviation vector, we first evaluate the ones involving the metric perturbations only. The first order change in HEE is,
\begin{gather}
\Delta^{(1)}S={1\over 4G_N}\Delta^{(1)}A={1\over 8G_N} \int d^{d-1}\tau~\sqrt{h}h^{ab}\accentset{(1)}{P}(\partial_{a},\partial_{b})
=\frac{2L}{32 G_N} \pi  z_*^2 (2 C+D)\notag\\=\frac{2L\times l^2}{4G_Nz_0^3}\frac{  (1+2\beta^2\gamma^2)\Gamma \left(\frac{1}{4}\right)^2}{32~ \Gamma \left(\frac{3}{4}\right)^2},
 \end{gather}
 which again matched with the results obtained in \cite{Mishra:2015cpa,Mishra:2016yor}. As before the last three terms in the second variation formula are,
 \begin{gather}
\int d^{d-1}\tau~\sqrt{h}{h^{ab}\over 2}\accentset{(2)}{P}(\partial_{a},\partial_{b})=\frac{2L\times\pi ^{3/2} c^5 (7 C'+5 D')}{21 \sqrt{2} \Gamma \left(\frac{3}{4}\right)^2}
\end{gather}
The next term which involves the product of perturbations is,
\begin{gather}
 \int d^{d-1}\tau~\sqrt{h}{1\over 2}h^{ac}h^{bd}\accentset{(1)}{P}(\partial_{a},\partial_{b})\accentset{(1)}{P}(\partial_{c},\partial_{d})=\frac{2L\times z_*^5 K\left(\frac{1}{2}\right) \left(77 C^2+45 D^2\right)}{231 \sqrt{2}}
 \end{gather}
 Finally we have the term
  \begin{gather}
 \int d^{d-1}\tau~\sqrt{h}{1\over 4}h^{ab}h^{cd}\accentset{(1)}{P}(\partial_{c},\partial_{d})\accentset{(1)}{P}(\partial_{a},\partial_{b})=\frac{2L\times\sqrt{\pi } z_*^5 \Gamma \left(\frac{5}{4}\right) \left(77 C^2+110 C D+45 D^2\right)}{462 \Gamma \left(\frac{3}{4}\right)}
 \end{gather}
Now we go over to the other integrals. Consider the term,
  \begin{gather}
\int d^{d-1}\tau~\sqrt{h} \Bigl(h^{ab}h^{cd}\accentset{(1)}{P}(\partial_b,\partial_d)g(N^{\perp},K(\partial_a,\partial_c))-h^{ab}g(C(\partial_a,\partial_b),N^{\perp})\Bigr)\notag\\
~~~~~~~~~~~~~~~~~~~~~~~~~~~~=\int_{-L}^{L}\int_{0}^{\pi}\frac{1}{2 z_*\sin ^{\frac{3}{2}}\theta}\bigg[ z_*^3 \left(\frac{3 C}{2}+3 D\right) \sin^{\frac{5}{2}} \theta-\frac{7}{2} z_*^3 D \sin ^{\frac{9}{2}}\theta\bigg]\Theta^1~d\theta~d\lambda
 \end{gather}
Note that $\Theta^1$ contains two terms. One that does not have an analytical form and the other which does. Lets write these as $\Theta^1=-\frac{\cos(\theta)}{\sqrt{2\sin(\theta)}}\int_{0}^{\theta}f(\theta')~d\theta'+G(\theta)+\Theta^1_c(\theta)$. Therefore the above integral becomes,
 \begin{eqnarray}
\int_{-L}^{L}\int_{0}^{\pi}&&\frac{1}{2 z_*\sin ^{\frac{3}{2}}\theta}\bigg[ z_*^3 \left(\frac{3 C}{2}+3 D\right) \sin^{\frac{5}{2}} \theta-\frac{7}{2} z_*^3 D \sin ^{\frac{9}{2}}\theta\bigg]\\\nonumber
&&~~~~~~~~~~~~~~~~~~~~~~~~~~~~\times\Bigg(-\frac{\cos(\theta)}{\sqrt{\sin(\theta)}}\int_{0}^{\theta}f(\theta')~d\theta'+G(\theta)+\Theta^1_c\Bigg)d\theta~d\lambda
\end{eqnarray}
 Note that the $G(\theta)$ can be obtained easily and the value evaluates to,
 \begin{gather}
\frac{2L\times\sqrt{\pi } z_*^5 \Gamma \left(\frac{9}{4}\right) \left(77 C^2+110 C D+29 D^2\right)}{352 \Gamma \left(\frac{11}{4}\right)}
 \end{gather}
  The complementary part of the solution gives,
 \begin{gather}
-\frac{2L}{64} \sqrt{\frac{\pi }{2}} z_*^5 \Gamma \left(\frac{1}{4}\right)^2 (2 C+D)^2
 \end{gather}
 The other integral is of the form $\int_{0}^{\pi} \left(g(\theta)\int_0^\theta f(\theta')d\theta'\right)d\theta$ and can be evaluated by parts,
 \begin{gather}
\int_{0}^{\pi}g(x)\int_{0}^{x}f(x')dx'dx=\left[\left(\int_{0}^{x}f(x')dx'\right)\left(\int g(x)dx\right)\right]_{0}^{\pi}-\int_{0}^{\pi}f(x)\int g(x')dx'dx
\end{gather}
The first term in the above expression does not contribute, while the second term reproduces the number obtained for $G(\theta)$.
%\begin{gather}
%-\frac{2l~\sqrt{\pi } z_{*}^5 \Gamma \left(\frac{13}{4}\right) \left(18865 C^2-75768 C D+21627 D^2\right)}{873180 \Gamma \left(\frac{11}{4}\right)}
%\end{gather}
%By a similar technique the final integral can be evaluated,
 %\begin{gather}
%-\int d^{d-1}\tau~\sqrt{h} \Bigl(h^{ab}g(C(\partial_a,\partial_b),N^{\perp})\Bigr)
% \end{gather}
%The $G(\theta)$ term in the above expression evaluates to,
%\begin{gather}
%\frac{2l~\sqrt{\pi } z_{*}^5 \Gamma \left(\frac{9}{4}\right) \left(3773 C^2-32802 C D+70461 D^2\right)}{155232 \Gamma \left(\frac{11}{4}\right)}
%\end{gather}
%The complementary part of the solution evaluates to, 
%\begin{gather}
%\frac{2l~z_{*}^5 (14 C-57 D) \left(\sqrt{2 \pi } \Gamma \left(\frac{1}{4}\right)^2 (57 D-14 C)+56 \pi  C-264 \pi  D\right)}{56448}
%\end{gather}
%%%%%%%%%%%%%%%%%%%%%%%%%%%%%%%%%%%%%%%%%%%%%%%%%%%%%%%%%%%%%%%%%%%%%%%%%%%%%%%%%%%%%%%%%%%%%%%%%%%%%%%%%%%%%%%%%%%%%%%%%%%%%%%%%%%%%%%%%%%%%%%%%%%%%%%%%%%%%%%%%%%%%%%%%%%%%%%%%%%%%%%%%%%%%%%%%%%%%%%%%%%%%%%%%%%%%%%%%%%%%%%%%%%%%%%%%%%%%%%%%%%%%%%%%%%%%%%%%%%%%%%%%%%%%%%%%%%%%%%%%%%%%%%%%%%%%%%%%%%%%%%%%%%%%%%%%%%%%%%%%%%%%%%%%%%%%%%%%%%%%%%%%%%%%%%%%%%%%%%%%%%%%%%%%%%%%%%%%%%%
%The final term by integration by parts gives
%\begin{gather}
%\frac{2l~\sqrt{\pi } z_{*}^5 \Gamma \left(\frac{9}{4}\right) \left(3773 C^2-32802 C D+70461 D^2\right)}{155232 \Gamma \left(\frac{11}{4}\right)}
%\end{gather}
The total variation $\Delta^{(2)}S$ is then given as,
\begin{gather}
\Delta^{(2)}S={1\over 4G_N} \Delta^{(2)}A=\frac{2L\times l^5 }{z_0^6}\frac{\Gamma \left(\frac{1}{4}\right)^5}{\Gamma \left(\frac{3}{4}\right)^7}\frac{ \left(-84 (\pi -1) \beta ^4 \gamma ^4+28 (4-3 \pi ) \beta ^2 \gamma ^2 +(48-21 \pi)\right)}{21504\times 4 G_N\sqrt{2} \pi}
\end{gather}
This expression gives the second order change of HEE. As in the case of circular disk, the positivity of relative entropy demands that $\Delta^{(2)}A\leq 0$. This can be checked through a plot of $\Delta^{(2)}A$ against $\beta$ (See Fig-1). The whole expression is negative (at $\beta=0$) and monotonically decreasing as $\beta$. The change $\Delta S$ or the plot cannot however be trusted for too large values of $\beta$, since one needs to add further higher order corrections to the change for large $\beta$.
\begin{figure}[h]
 \centering
 \begin{overpic}[width=10cm]{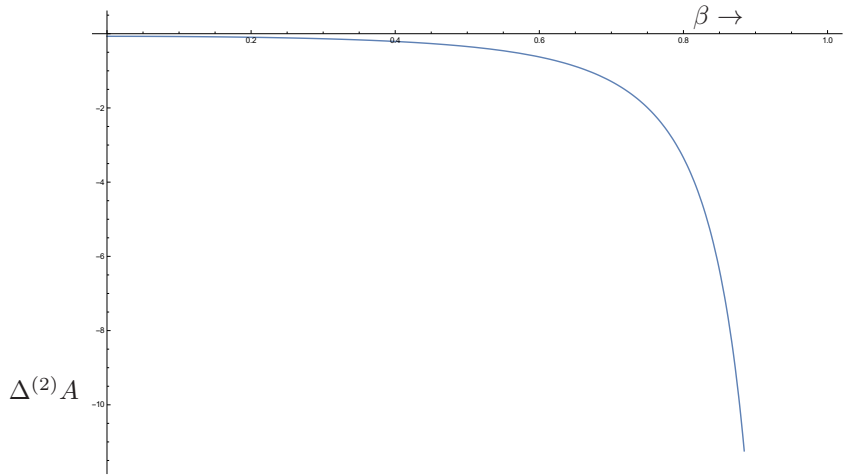}
\put(-11,10){\small$\Delta^{(2)}A$}
%\put(80,22){\small$S_{th}=\pi^2$}
%\put(80,41){\small$S_{th}=2\pi^2$}
\put(80,60){\small$\beta\rightarrow$}
%\put(105,0){\Large$l$}
\end{overpic}
 \caption{\it Plot of $\Delta^{(2)}A~~\text{vs}~~\beta$ for strip along $x$ boost along $y$}
\end{figure}

The full expression for change of HEE is then given by 
\begin{gather}
 \Delta S=\Delta^{(1)}S+\frac{1}{2}\Delta^{(2)}S\notag\\
=\frac{2L\times l^2}{4G_N z_0^3}\frac{  (1+2\beta^2\gamma^2)\Gamma \left(\frac{1}{4}\right)^2}{32~ \Gamma \left(\frac{3}{4}\right)^2}+\frac{2L\times l^5 }{2\times z_0^6}\frac{\Gamma \left(\frac{1}{4}\right)^5}{\Gamma \left(\frac{3}{4}\right)^7}\frac{ \left(-84 (\pi -1) \beta ^4 \gamma ^4+28 (4-3 \pi ) \beta ^2 \gamma ^2+(48-21 \pi) \right)}{4 G_N\times 21504~\sqrt{2} \pi}
\end{gather}
the above expression gives the net change in HEE for strip entangling surface upto second order over pure $AdS$(ground state) value.
%%%%%%%%%%%%%%%%%%%%%%%%%%%%%%%%%%%%%%%%%%%%%%%%%%%%%%%%%%%%%%%%%%%%%%%%
%%%%%%%%%%%%%%%%%%%%%%%%%%%%%%%%%%%%%%%%%%%%%%%%%%%%%%%%%%%%%%%%%%%%%%
%%%%%%%%%%%%%%%%%%%%%%%%%%%%%%%%%%%%%%%%%%%%%%%%%%%%%%%%%%%%%%%%%%%%%%%%%%
%%%%%%%%%%%%%%%%%%%%%%%%%%%%%%%%%%%%%%%%%%%%%%%%%%%%%%%%%%%%%%%%%%%%%%%%%%%%
%%%%%%%%%%%%%%%%%%%%%%%%%%%%%%%%%%%%%%%%%%%%%%%%%%%%%%%%%%%%%%%%%%%%%%%%%%%%%
 \subsubsection{\underline{Strip along `x' boost along `y'}}
 In this case all the integrals for $e_1$ are same as that of the previous case with $C,~D$ replaced by $\tilde C,~\tilde D$ and $C',~D'$ replaced by $\tilde C',~\tilde D'$ (see appendix \ref{Pert}). However in this case the non homogeneous part of the $e_0$ equation is non trivial. Hence the extremal surface doesn't remain on the same time slice . The equation is,
 \begin{gather}
4\sin^2{\theta}~\partial^2_\theta\alpha^0+2\sin{\theta}\cos{\theta}~\partial_{\theta}\alpha^0+{z_*}^2\sin{\theta}~\partial^2_\lambda\alpha^0-2\alpha^0=-3z_*^3(\sin{\theta})^{\frac{5}{2}}B\cos{\theta},
\end{gather}
which following the previous arguments reduces to solving only the equation, 
 \begin{gather}
{d^2\Theta^0\over d\theta^2}+\frac{1}{2}\cot{\theta}{d\Theta^0\over d\theta}-\frac{1}{2}\cosec^2\theta\Theta^0=-\frac{3}{4}z_*^3(\sin{\theta})^{\frac{1}{2}}B\cos{\theta}
\end{gather}
The solutions of this can be obtained in a straightforward manner and therefore we do not have to resort to efforts made in the previous section. %The homogeneous solution turns out to be of the form,
%\begin{gather}
%\Theta^0_c(\theta)=-\frac{2 C_1 E\bigg(\left.\frac{1}{4} (\pi -2 \theta)\right|2\bigg)}{\sqrt{\sin (\theta)}}+\frac{C_2}{\sqrt{\sin (\theta)}}
%\end{gather}
The full solutions turns out to be of the form,
\begin{gather}
\Theta^0=\frac{-\tilde B z_*^3 \theta}{4\sqrt{\sin\theta}}+\frac{\tilde B z_*^3 \sin2\theta}{8\sqrt{\sin\theta}}-\frac{2 C_1 E\bigg(\left.\frac{1}{4} (\pi -2 \theta)\right|2\bigg)}{\sqrt{\sin (\theta)}}+\frac{C_2}{\sqrt{\sin (\theta)}}
\end{gather}
Imposing the conditions $\Theta=0$ at $\theta=0$ and $\theta=\pi$, fixes $C_1$ and $C_2$ to,
%\begin{gather}
%\sqrt{2} C_1 \left(K\left(\frac{1}{2}\right)-2 E\left(\frac{1}{2}\right)\right)+C_2=0\\
%\pi  \tilde B z_*^3-4 \left(\sqrt{2} C_1 \left(2 E\left(\frac{1}{2}\right)-K%\left(\frac{1}{2}\right)\right)+C_2\right)=0,
%\end{gather}
the solutions of which are,
\begin{gather}
C_1=\frac{\pi  \tilde B z_*^3}{8 \sqrt{2} \left(2 E\left(\frac{1}{2}\right)-K\left(\frac{1}{2}\right)\right)},~~C_2=\frac{1}{8} \pi  \tilde B z_*^3,
\end{gather}
where $K(\alpha)$ and $E(\alpha)$ are the complete elliptic integral of the first and second kind respectively. The contributions coming from the component $\alpha^1$ of the deviation vector turns out to be same as that in the previous section with $C,~D$ replaced by $\tilde C,~\tilde D$ and $C',~D'$ replaced by $\tilde C',~\tilde D'$. The only other contribution different from the previous case comes from $-Tr(C)$ for the component $\alpha^0$ of the deviation vector and evaluates to,
\begin{gather}
\frac{2L ~\pi ^{3/2} (21 \pi -80) B^2 z_{*}^5}{336 \sqrt{2} \Gamma \left(\frac{3}{4}\right)^2}
\end{gather}
Total variation $\Delta^{(2)}S$ without the previous term is then given by,
\begin{gather}
\Delta^{(2)}S=\frac{2L\times l^5 }{4 G_N z_0^6}~~\frac{\Gamma \left(\frac{1}{4}\right)^5}{\Gamma \left(\frac{3}{4}\right)^7}\left(\frac{(20-21 \pi ) \beta ^4 \gamma ^4+2 (40-21 \pi ) \beta ^2 \gamma ^2+2 (21 \pi -80) \beta  \gamma ^4+(48-21 \pi) }{21504 \sqrt{2} \pi }\right)
\end{gather}
 As in the previous case $\Delta^{(2)}A\leq 0$. This can be checked by plotting $\Delta^{(2)}S$ against $\beta$(see Fig-2). It is negative and monotonically decreasing as a function of $\beta$. It is important to note that the boost independent term in the expression for $\Delta^{(2)}S$ for both the cases is same. Setting boost to zero makes both the cases identical to AdS black brane geometry.
\begin{figure}[h]
 \centering
 \begin{overpic}[width=10cm]{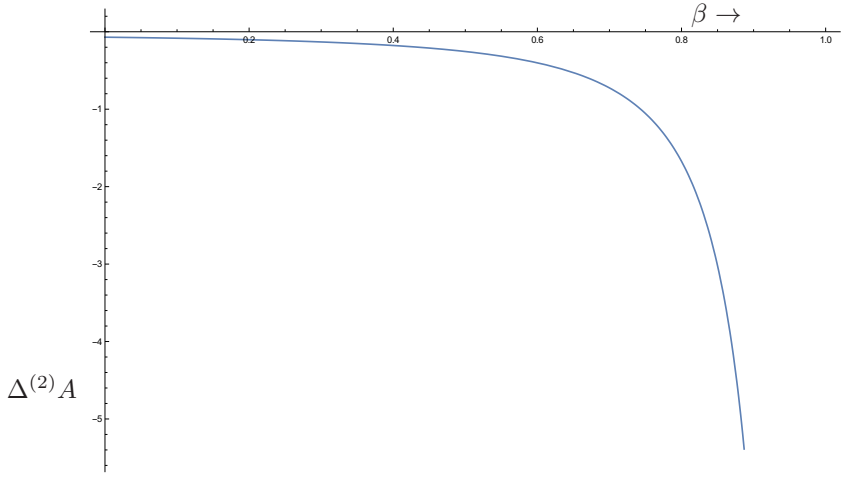}
\put(-11,10){\small$\Delta^{(2)}A$}
%\put(80,22){\small$S_{th}=\pi^2$}
%\put(80,41){\small$S_{th}=2\pi^2$}
\put(80,60){\small$\beta\rightarrow$}
%\put(105,0){\Large$l$}
\end{overpic}
 \caption{\it Plot of $\Delta^{(2)}A~~\text{vs}~~\beta$ for strip along $x$ boost along $y$}
\end{figure}

The first order change in HEE is given by
\begin{gather}
\Delta^{(1)}S={1\over 4G_N}\Delta^{(1)}A=\frac{2L\times l^2}{4 G_N z_0^3}\frac{  (1+\beta^2\gamma^2)\Gamma \left(\frac{1}{4}\right)^2}{32~ \Gamma \left(\frac{3}{4}\right)^2}
\end{gather}
Thus the full expression for change in HEE is then given by 
\begin{gather}
 \Delta S=\Delta^{(1)}S+\frac{1}{2}\Delta^{(2)}S\notag\\
 =\frac{2L\times l^2}{4 G_N z_0^3}\frac{  (1+\beta^2\gamma^2)\Gamma \left(\frac{1}{4}\right)^2}{32~ \Gamma \left(\frac{3}{4}\right)^2}\notag\\+\frac{2L\times l^5 }{8 G_N z_0^6}~~\frac{\Gamma \left(\frac{1}{4}\right)^5}{\Gamma \left(\frac{3}{4}\right)^7}\left(\frac{(20-21 \pi ) \beta ^4 \gamma ^4+2 (40-21 \pi ) \beta ^2 \gamma ^2+2 (21 \pi -80) \beta  \gamma ^4+(48-21 \pi) }{21504 \sqrt{2} \pi }\right)
\end{gather}
the above expression gives the net change in HEE for strip entangling surface upto second order over pure $AdS$(ground state) value.
%\begin{gather}
%\frac{\pi  c^5 \left(-231 \pi ^{3/2} \sqrt{2} \left(117 \beta ^2 \gamma ^2+53\right) \left(171 \beta ^2 \gamma ^2+43\right)+4 \sqrt{2 \pi } \left(675 \beta ^2 \gamma ^2 \left(5043 \beta ^2 \gamma ^2+3670\right)+710018\right)+924 \Gamma \left(\frac{3}{4}\right)^2 \left(81 \beta ^2 \gamma ^2+34\right) \left(171 \beta ^2 \gamma ^2+43\right)\right)}{104315904 \Gamma \left(\frac{7}{4}\right)^2}+\frac{\pi ^{3/2} (21 \pi -80) \beta ^2 \gamma ^4 c^5}{336 \sqrt{2} \Gamma \left(\frac{3}{4}\right)^2}
%\end{gather}
\section{Issues of Gauge dependence}
The $\Phi_\lambda$'s in section \ref{DOJE} are called the identification maps. It encodes the information about how points in the perturbed and the unperturbed space times are to be identified. The notion of gauge transformation can be shown to arise due to different choices of the $\Phi_\lambda$'s. It is evident that the identification maps can be so chosen that the location of the perturbed minimal surface in the unperturbed spacetime is same as that of the unperturbed minimal surface. This is precisely the interpretation of the Hollands-Wald gauge \cite{Hollands:2012sf} used in \cite{Lashkari:2015hha,Faulkner:2017tkh,Beach:2016ocq}. But it seems that this in general can be done at any order of perturbation and not just at the linear order. Further, it seems that by choice of such gauge one renders the inhomogeneous term, in the Jacobi equation obtained, trivial and therefore irrelevant. We must emphasize that this is not the case. In order to find the Hollands Wald gauge (at linear order) one has to solve a linear second order differential equation which is precisely the inhomogeneous Jacobi equation. This has also been pointed out in \cite{Mosk:2017vsz}. Therefore choosing the Hollands-Wald gauge does not trivialize the problem of finding the change in area. However, it is absolutely possible that the Holland- Wald gauge is a convenient choice if one tries to find identities that the higher order perturbations of the area functional satisfy or finding relations between two gauge independent quantities like the `Fisher information` and the canonical energy \cite{Lashkari:2015hha}.

Having discussed this it is quite viable to state that the inhomogeneous equation is gauge covariant. In other words any gauge transformation of the metric perturbation can be absorbed in a shift of the deviation vector itself. This is a quite plausible conclusion that follows from the following lemma due to \cite{Stewart:1974uz}. The linear perturbation $Q_1$ of a quantity $Q_0$ on $(M, g)$ is gauge invariant if and only if one of the following holds:
(i) $Q_0$ vanishes,
(ii) $Q_0$ is a constant scalar,
(iii) $Q_0$ is a constant linear combination of products of Kronecker deltas.  
 In our case $Q_0$ is the mean curvature $(H)$ of the extremal surface in the background spacetime and hence is identically zero. However there is a subtle issue in application of the above lemma in our case. The quantities $Q$ defined in the lemma are globally defined while $H$ is locally defined on a codimension two surface. The expression for the second variation of the area functional is however invariant under different choices of $\Phi_{\lambda}:M\rightarrow M_{\lambda}$.

\section{Conclusion}
A few comments about higher order perturbations are in order. As is usual with any perturbation theory, the homogeneous part of the second order perturbation equation would be same as the Jacobi equation. However the inhomogeneous term will now depend both on second order perturbations as well as first order deviations. Note the second order deviation vector $M$ (say), can always be taken to commute with $N$ owing to the fact that they represent independent variations. Since the normal bundle is two dimensional one can have at most two mutually commuting directions. Hence it seems that the perturbation will terminate at second order and the complete change of entanglement entropy can be obtained by exponentiating this change upto second order. However this is speculative and requires further investigation. We have presented a systematic approach to obtain the change in HEE up to second order. For simplicity we have calculated this in $4$-dimensions but the approach remains unchanged in higher dimensions. The inhomogeneous Jacobi equation and second variation of the area functional presented here can be applied to non $AdS$ geometries also. In fact the Jacobi operator simplifies for the asymptotically flat case. We have seen that second order change receives contributions from first order changes in the embeddings and second order change of the bulk metric. In this approach the nature of the flow of the extremal surface can be understood by looking at the components of the deviation vector. Further, having obtained the second variation one can check if more general entropy bounds \cite{Bekenstein:1980jp,Hod:2000ju,Bekenstein:1999cf,Park:2015hcz} or has any relation with geometric inequalities \cite{Dain:2011mv} in general.

Note added: While this manuscript was being completed, a relevant work \cite{Mosk:2017vsz} which takes an iterative approach to calculate the higher order change in HEE, came up. The expression used for second variation of area evaluate is similar to that obtained by us. A specific choice of gauge (Holland-Wald gauge) was made to obtain certain results for spherical subsystem. However in our work we use the Fefferman Graham gauge and explicitly solve the inhomogeneous equation for both spherical and strip subsystems, with Boosted Black Brane like perturbations and evaluate the area variation upto second order. 

\section{acknowledgements}
The authors would like to thank Amit Ghosh, Lucas Ambrozio and Harvendra Singh for discussions. AG acknowledges Sudipta Sarkar for helpful discussions and for pointing out reference \cite{Mosk:2017vsz} on the very day it appeared on arXiv. AG also acknowledges Atul Dixit for his course on special functions and the hospitality at SINP, where part of this work was done . Most of the integrals and solutions of the differential equations were done using Maple \cite{maple} and Mathematica \cite{mathematica}. Tensor calculations were done with the GRTensor package \cite{grtensor} on a Maple platform. AG is supported by SERB, government of India through the NPDF grant (PDF/2017/000533).

\appendix
\section{Revisiting the derivation of the inhomogeneous Jacobi equation for geodesics}\label{geosection}
To make sure that in the used notation the equation we have obtained is indeed the correct equation we are looking for, we will derive the inhomogeneous Jacobi equation derived in \cite{Ghosh:2016fop}.

Note that the geodesic equation can be written as $(\nabla_{T}T)^\perp=0$ (where $T$ is the tangent vector to the geodesic and satisfies $\nabla_TT=fT$). We will consider a variation of the geodesic under  $\delta_N$. The variation accounts for both change of embeddings and metric perturbations.
\begin{gather}\label{geo}
 \delta_{N}(\nabla_{T}T)^\perp=\delta_{N}(\nabla_{T}T)-\delta_{N}(\nabla_{T}T)^T\\\notag%=\nabla_{N}(\nabla_{T}T)+\delta_{P}(\nabla_{T}T)-\nabla_{N}(\nabla_{T}T)^T-\delta_{P}(\nabla_{T}T)^T\\\notag
 =\nabla_{T}^2 N+R(N,T)T+C(T,T)-\nabla_N (fT)-\delta_{g}(f)T
\end{gather}
Our convention implies that $\delta_P (\nabla_X Y)=C(X,Y)$. Noting that $f={g(\nabla_T T,T)\over g(T,T)}$, we can find the variation $\delta_g f$. After a few algebraic steps one gets the following expression,
\begin{gather}\label{deltagf}
% (\delta_P{f})g(T,T)+f\delta_P(g(T,T))=\delta_P(g(\nabla_T T,T))\\\notag
 %\Rightarrow(\delta_P{f})g(T,T)+f P(T,T)=P(\nabla_T T,T)+g(C(T,T),T)\\\notag
 %\Rightarrow(\delta_P{f})g(T,T)+\cancel{f P(T,T)}=\cancel{f P(T,T)}+g(C(T,T),T)\\\notag
 %\Rightarrow
 \delta_P{f}={g(C(T,T),T)\over g(T,T)}
\end{gather}
Also note that,
\begin{gather}\label{gradNf}
 %(\nabla_T N)^T =\nabla_{T}^2 N-(\nabla_{T}^2 N)^T+R(N^\perp,T)T-(f\nabla_TN-f(\nabla_TN)^T)\\
%+C(T,T)^\perp
 %f=\frac{g(\nabla_TT,T)}{g(T,T)}\\
 %\implies 
 \nabla_Nf=\frac{g(\nabla_T^2N,T)}{g(T,T)}-\frac{fg(\nabla_TN,T)}{g(T,T)}
\end{gather}

Substituting \eqref{deltagf} and \eqref{gradNf} in \eqref{geo} we get
\begin{gather}\label{geo1}
 \delta_{N}(\nabla_{T}T)^\perp=\nabla_{T}^2 N-(\nabla_{T}^2 N)^T+R(N^\perp,T)T-(f\nabla_TN-f(\nabla_TN)^T)+C(T,T)^\perp.
\end{gather}
Equating the above to zero gives the inhomogeneous equation,
 \begin{gather}\label{geo1}
 \nabla_{T}^2 N^\perp+R(N^\perp,T)T-f\nabla_TN^\perp+C(T,T)^\perp=0.
\end{gather}
In \cite{Ghosh:2016fop} the unperturbed geodesic was taken to be affinely parametrised. Therefore putting $f=0$ in the above equation reproduces the equation obtained.
 %It should be noted that $N$ has both tangent and normal part. Thus the tangent part should also cancel out from \cref{geo1}.
 %\begin{gather*}
  %\nabla_{T}^2 N^T+R(N^T ,T)T-\nabla_{N^T} (fT)=0
 %\end{gather*}
%I have not been able to derive this. I add here a few of the results that I have obtained
%\begin{gather*}
 %(\nabla_T N)^T =\nabla_T N^T,~ (\nabla_T N)^\perp =\nabla_T N^\perp,~(\nabla_T^2 N)^T=\nabla_T^2 N^T,\\
 %f=\frac{g(\nabla_TT,T)}{g(T,T)}\\
 %\implies \nabla_Nf=\frac{g(\nabla_T^2N,T)}{g(T,T)}-\frac{fg(\nabla_TN,T)}{g(T,T)}
%\end{gather*}
%Replacing this into \cref{geo1} gives
%\begin{gather}\label{geo2}
%\nabla_{T}^2 N+R(N,T)T-\nabla_N f~T-f\nabla_TN+C(T,T)^\perp=0\\
%\implies\nabla_{T}^2 N-(\nabla_{T}^2 N)^T+R(N^\perp,T)T-(f\nabla_TN-f(\nabla_TN)^T)\\
%+C(T,T)^\perp=0
%\end{gather}
%Note that $R(N^T,T)T=0$
%\end{tcolorbox}
 \section{Boosted Black brane as a perturbation over AdS}\label{Pert}
 
 The boosted black brane metric in holographic coordinates is of the following form 
\begin{gather}
 ds^2={R^2\over z^2}\left[-\mathcal {A}(z)dt^2+\mathcal{B}(z)dx^2+\mathcal{C}(z)dtdx+dx^2+{dz^2\over f(z)}\right],
\end{gather}
where,
\begin{gather*}
 \mathcal{A}(z)=1-\gamma^2({z\over z_0})^3,~~\mathcal{B}(z)=1+\beta^2\gamma^2({z\over z_0})^3,\\
 \mathcal{C}(z)=2\beta\gamma^2({z\over z_0})^3,~~f(z)=1-({z\over z_0})^3
\end{gather*}
 $z_0$ is the location of the horizon and $0\leq\beta\leq1$ is the boost parameter, while $\gamma={1\over\sqrt{1-\beta^2}}$. With the boost along $x$ direction. The boosted black brane is a finite change from $AdS$ and hence cannot be observed as a perturbation over it.  In order to see it as a perturbation over $AdS$, we have to write it in suitable asymptotic (Fefferman Graham)  coordinates. The Fefferman Graham coordinates are obtained by demanding \cite{Skenderis:1999nb, deHaro:2000vlm}
\begin{gather}
 {dz\over z\sqrt{f(z)}}={d\rho\over\rho}
\end{gather}
Integrating this and setting the integration constant to $({\rho_0}^3=4{z_0}^3)$ we get
\begin{gather}
 {1\over z^2}={1\over\rho^2}(1+({\rho\over{\rho_0}})^3)^{4\over 3}={1\over\rho^2}g(\rho)^{4\over 3}
\end{gather}
Now we expand the metric coefficient upto second order in $({\rho\over{\rho_0}})^3$, Substituting this back in the metric we get
\begin{gather}
 ds^2
 ={R^2\over\rho^2}\Biggl[d\rho^2+\left(\eta_{\mu\nu}+\rho^3\gamma^{(3)}_{\mu\nu}+\rho^6\gamma^{(6)}_{\mu\nu}\right)dx^{\mu}dx^{\nu}\Biggr]
\end{gather}
Where
\begin{gather}\label{got}
 \gamma^{(3)}_{\mu\nu}=\begin{bmatrix}
        -({1\over 3}-\gamma^2)({1\over z_0})^3 & \beta\gamma^2({1\over z_0})^3 & 0 \\
          \beta\gamma^2({1\over z_0})^3 & \left({1\over 3}+\beta^2\gamma^2\right)({1\over z_0})^3 & 0\\
          0 & 0 & {1\over 3}({1\over z_0})^3
        \end{bmatrix}
\end{gather}
One can check that $Tr( \gamma^{(3)}_{\mu\nu})=0$ and
\begin{gather}
 \gamma^{(6)}_{\mu\nu}=\begin{bmatrix}
        -\left({2\over 9}+{8\over 3}\gamma^2\right){1\over16 {z_0}^6} & -{1\over 6}\beta\gamma^2({1\over z_0})^6 & 0 \\
         -{1\over 6}\beta\gamma^2({1\over z_0})^6  & \left({2\over 9}-{8\over 3}\beta^2\gamma^2\right){1\over 16 {z_0}^6} & 0\\
          0 & 0 & {2\over 9}{1\over 16{z_0}^6}
        \end{bmatrix}
\end{gather}
The perturbation $\accentset{(1)}{P}_{\mu\nu}$  and $\accentset{(2)}{P}_{\mu\nu}$ can be read off as, $\accentset{(1)}{P}_{\mu\nu}=\gamma^{(3)}_{\mu\nu}z$ and $\frac{1}{2}\accentset{(2)}{P}_{\mu\nu}=\gamma^{(6)}_{\mu\nu}z^4$ respectively. To calculate the non homogeneous term in the Jacobi equation, we need the expression for $C(\partial_\mu,\partial_\nu)$, which in a given coordinate system can be written as, 
\begin{gather}\label{oye}
{C}^{\mu}_{\nu \rho}(x)=\frac{1}{2}{g}^{\mu\sigma}\left(\partial_\nu \accentset{(1)}{P}_{\rho\sigma}+\partial_\rho \accentset{(1)}{P}_{\nu \sigma}-\partial_\sigma \accentset{(1)}{P}_{\nu \rho}\right)-\frac{1}{2}\accentset{(1)}{P}^{\mu \sigma}\left(\partial_\nu{g}_{\rho \sigma}+\partial_\rho{g}_{\nu \sigma}-\partial_\sigma{g}_{\nu \rho}\right)
\end{gather}
Note that this quantity is a vector field in the tangent bundle and therefore it's coordinate expression has three indices. 
We will calculate this for boosts both in the $x$ direction and the $y$ direction. Note that though the direction of the boost does not affect the results for a spherical boundary subsystem, it does so for the strip subsystem. In the Fefferman graham gauge the expression for $C(\partial_\mu,\partial_\nu)$. 

For boost along the $x$ axis, the expression for $\accentset{(1)}{P}(\partial_\mu,\partial_\nu)$ and $\accentset{(2)}{P}(\partial_\mu,\partial_\nu)$ is of the following form.

{\[\displaystyle \accentset{(1)}{P} _{\mu\nu}=\left(\begin{array}{cccc}A ~z  & B ~z  & 0 & 0 \\B ~z  & C ~z  & 0 & 0 \\0 & 0 &  D~z  & 0 \\0 & 0 & 0 & 0 \\\end{array}\right)~~~~~~~\displaystyle \frac{1}{2}\accentset{(2)}{P} _{\mu\nu}=\left(\begin{array}{cccc}A' ~z^4  & B' ~z^4  & 0 & 0 \\B' ~z^4  & C' ~z^4  & 0 & 0 \\0 & 0 &  D'~z^4  & 0 \\0 & 0 & 0 & 0 \\\end{array}\right)\]}
The quantity $C^\mu_{\nu\rho}$  can be calculated from eqn (\ref{oye}),
\begin{gather} 
C ^{z }~_{t ~t }=-\frac{1}{2}~z ^{2}~A, ~~C ^{z }~_{x ~t }=-\frac{1}{2}~z ^{2}~B, ~~  C ^{t }~_{z ~t }=-\frac{3}{2}~z ^{2}~A, ~~ C ^{x }~_{z ~t }=\frac{3}{2}~z ^{2}~B\notag\\
C ^{z }~_{t ~x }=-\frac{1}{2}~z ^{2}~B,~~\displaystyle C ^{z }~_{x ~x }=-\frac{1}{2}~z ^{2}~C,~~ C ^{t }~_{z ~x }=-\frac{3}{2}~z ^{2}~B,~~C ^{x }~_{z ~x }=\frac{3}{2}~z ^{2}~C\notag\\
 C ^{z }~_{y ~y }=-\frac{1}{2}~z ^{2}~D,~~ C ^{y }~_{z ~y }=\frac{3}{2}~z ^{2}~D,~~ C ^{t }~_{t ~z }=-\frac{3}{2}~z ^{2}~A,~~ C ^{x }~_{t ~z }=\frac{3}{2}~z ^{2}~B\notag\\
C ^{t }~_{x ~z }=-\frac{3}{2}~z ^{2}~B,~~C ^{x }~_{x ~z }=\frac{3}{2}~z ^{2}~C,~~C ^{y }~_{y ~z }=\frac{3}{2}~z ^{2}~D,
\end{gather}

where $C,D$ can be read off from the previous expression for $P$'s and $\gamma$'s eqn. (\ref{got}) and is given as $C=\left({1\over 3}+\beta^2\gamma^2\right){1\over z_0^3},~D={1\over 3}{1\over z_0^3}$. The components of $\frac{1}{2}\accentset{(2)}{P}_{\mu\nu}$ will be $C^\prime,D^\prime$ and is given as $C'=\left({2\over 9}-{8\over 3}\beta^2\gamma^2\right){1\over 16 {z_0}^6},~D'={2\over 9}{1\over 16 {z_0}^6}$. 

For boost along the $y$ axis, $\accentset{(1)}{P}(\partial_\mu,\partial_\nu)$ and $\accentset{(2)}{P}(\partial_\mu,\partial_\nu)$ is of the form,

{\[\displaystyle \accentset{(1)}{P} _{\mu\nu}=\left(\begin{array}{cccc}\tilde A ~z  & 0 & \tilde B ~z  & 0 \\0 & \tilde C ~z  & 0 & 0 \\\tilde B ~z  & 0 & \tilde D~z  & 0 \\0 & 0 & 0 & 0 \\\end{array}\right)~~~~\displaystyle \frac{1}{2}\accentset{(2)}{P} _{\mu\nu}=\left(\begin{array}{cccc}\tilde A' ~z^4  & 0 & \tilde B' ~z^4  & 0 \\0 & \tilde C' ~z^4  & 0 & 0 \\\tilde B' ~z^4  & 0 & \tilde D'~z^4  & 0 \\0 & 0 & 0 & 0 \\\end{array}\right)\]}\\
 
 The quantity $C^\mu_{\nu\rho}$ is therefore,
 \begin{gather}
  C ^{z }~_{t ~t }=-\frac{1}{2}~z ^{2}~\tilde A,~~C ^{z }~_{y ~t }=-\frac{1}{2}~z ^{2}~\tilde B,~~C ^{t }~_{z ~t }=-\frac{3}{2}~z ^{2}~\tilde A,~~ C ^{y }~_{z ~t }=\frac{3}{2}~z ^{2}~\tilde B\notag\\
   C ^{z }~_{x ~x }=-\frac{1}{2}~z ^{2}~\tilde C,~~C ^{x }~_{z ~x }=\frac{3}{2}~z ^{2}~\tilde C,~~C ^{z }~_{t ~y }=-\frac{1}{2}~z ^{2}~\tilde B,~~C ^{z }~_{y ~y }=-\frac{1}{2}~z ^{2}~\tilde D\notag\\
   C ^{t }~_{z ~y }=-\frac{3}{2}~z ^{2}~\tilde B,~~ C ^{y }~_{z ~y }=\frac{3}{2}~z ^{2}~\tilde  D,~~C ^{t }~_{t ~z }=-\frac{3}{2}~z ^{2}~\tilde A,~~C ^{y }~_{t ~z }=\frac{3}{2}~z ^{2}~\tilde B\notag\\
    C ^{x }~_{x ~z }=\frac{3}{2}~z ^{2}~\tilde C,~~C ^{t }~_{y ~z }=-\frac{3}{2}~z ^{2}~\tilde B,~~\displaystyle C ^{y }~_{y ~z }=\frac{3}{2}~z ^{2}~\tilde D
 \end{gather}
where $\tilde{C}={1\over 3}({1\over z_0})^3,~~\tilde{D}=\left({1\over 3}+\beta^2\gamma^2\right)({1\over z_0})^3,~~ 
 \tilde{C}^\prime={2\over 9}{1\over 16{z_0}^6},~~\tilde{D}^\prime=\left({2\over 9}-{8\over 3}\beta^2\gamma^2\right){1\over 16 {z_0}^6},~~B=\tilde{B}=\beta\gamma^2({1\over z_0})^3$. This completes our first step in calculation of area, now we can proceed with solving the inhomogeneous Jacobi equation. 

\end{document}